\documentclass[]{svmult}

 \usepackage{graphicx}
 
\usepackage{makeidx}
\makeindex
 
 \usepackage{multicol}
 \usepackage{footmisc}

 \usepackage{bm}
 \usepackage{amsmath}
 \usepackage{amssymb}
 \usepackage{dsfont}
 \usepackage{latexsym}
 \usepackage{amsfonts}
 \usepackage{epsfig}
 \usepackage{epstopdf}
 \usepackage{color}
 \definecolor{darkblue}{rgb}{0,0,.5}
 \usepackage[linktocpage, colorlinks=true, linkcolor=darkblue, citecolor=darkblue]{hyperref}
 \usepackage[all]{hypcap}
 \usepackage[noadjust]{cite}

  \usepackage{mathrsfs}
 \usepackage{graphicx}
\usepackage{dcolumn}
\usepackage{bbm}
\usepackage{bm}
\usepackage{overpic}
 
  \newcommand{\ket}[1]{\left|#1\right>}
  \newcommand{\bra}[1]{\left<#1\right|}
  \newcommand{\expval}[1]{\left< #1 \right>}

\begin{document}

\title*{Chimera states in quantum mechanics}

\author{V. M. Bastidas, I. Omelchenko, A. Zakharova, E. Sch\"oll, and T. Brandes}
\institute{Institut f\"ur Theoretische Physik, Technische Universit\"at Berlin, Hardenbergstr. 36, D-10623 Berlin, Germany}
\maketitle

\begin{abstract}
 Classical chimera states are paradigmatic examples of \emph{partial synchronization patterns} emerging in nonlinear dynamics. These states are characterized by the spatial coexistence of two dramatically different dynamical behaviors, i.e., synchronized and desynchronized dynamics. Our aim in this contribution is to discuss signatures of chimera states in quantum mechanics.
 We study a network with a ring topology consisting of $N$ coupled quantum Van der Pol oscillators. We describe the emergence of chimera-like quantum correlations in the covariance matrix. Further, we establish the connection of chimera states to quantum information theory by describing the quantum mutual information for a bipartite state of the network. 
\end{abstract}

\section{Introduction}
%
\index{Chimera states, classical}
\index{Chimera states, quantum}
\index{Synchronization, classical}
\index{Synchronization, quantum}
\index{Phase oscillators}
\index{Quantum mutual information}
\index{R{\'e}nyi entropy}
%
Self-organization is one the most intriguing phenomenon in nature. Currently, there is a plethora of studies concerning pattern formation and the emergence of spiral waves, Turing structures, synchronization patterns, etc. In classical systems of coupled nonlinear oscillators, the phenomenon of chimera states, which describes the spontaneous emergence
of coexisting synchronized and desynchronized dynamics in networks of identical elements, has recently aroused much interest~\cite{PAN15}.  These intriguing spatio-temporal patterns were originally discovered in models of coupled phase oscillators~\cite{KUR02a,ABR04}. 
The last decade has seen an increasing interest in chimera states in dynamical
networks\cite{LAI09,MOT10,OME10a,OME12a,MAR10b,WOL11a,BOU14,LAI10,OME15}.  It was shown that they are not limited to
phase oscillators, but can be found in a large variety of different systems including time-discrete maps~\cite{OME11}, 
time-continuous chaotic models~\cite{OME12}, neural systems~\cite{OME13,HIZ13,VUE14a}, Van der Pol oscillators \cite{OME15a}, and Boolean networks~\cite{ROS14a}. 

Chimera states were found also in systems with higher spatial dimensions~\cite{OME12a,SHI04,MAR10,PAN13,PAN15a}.
New types of these peculiar states having multiple incoherent regions~\cite{SET08,OME13,MAI14,XIE14,VUE14a}, as well as amplitude-mediated~\cite{SET13,SET14}, and pure amplitude chimera states~\cite{ZAK14} were discovered.

The nonlocal coupling has been usually considered as a necessary condition for chimera states to evolve in systems of coupled oscillators. Recent
studies have shown that even global all-to-all coupling~\cite{SET14,YEL14,SCH14g,BOE15}, as well as more complex coupling topologies allow for the existence of chimera states~\cite{KO08,SHA10,LAI12,YAO13,ZHU14,OME15}. Furthermore,
time-varying network structures can give rise to alternating chimera states \cite{BUS15}.

Possible applications of chimera states in natural and technological systems include the phenomenon of unihemispheric sleep~\cite{RAT00}, bump states in neural systems~\cite{LAI01,SAK06a}, epileptic seizure~\cite{ROT14}, power grids~\cite{FIL08}, or social systems~\cite{GON14}.  
Many works considering chimera states have been based on numerical results. A deeper bifurcation analysis~\cite{OME13a} and even a possibility to control chimera states~\cite{SIE14c,BIC14} were obtained only recently.

The experimental verification of chimera states was first demonstrated in optical~\cite{HAG12} and chemical~\cite{TIN12,NKO13} systems. Further experiments involved  mechanical~\cite{MAR13}, electronic~\cite{LAR13,GAM14} and electrochemical~\cite{SCH14a, WIC13} oscillator systems, Boolean networks~\cite{ROS14a}, the optical comb generated by a passively mode-locked quantum dot laser~\cite{VIK14}, and superconducting quantum interference devices~\cite{LAZ15}.
 
While synchronization of classical oscillators has been well studied since the early observations of Huygens in the $17$th century~\cite{KAP12}, synchronization in quantum mechanics has only very recently become a focus of interest. For example, quantum signatures of synchronization in a network of globally coupled Van der Pol oscillators  have been investigated~\cite{SadeghpourPhysRevLett.111.234101,BruderPhysRevLett.112.094102}. Related works focus on the dynamical phase transitions of a network of nanomechanical oscillators with arbitrary topologies characterized by a coordination number~\cite{MarquardtPhysRevLett.111.073603}, and the semiclassical quantization of the Kuramoto model by using path integral methods~\cite{PachonKuramoto}.  

Contrary to classical mechanics, in quantum mechanics the notion of phase-space trajectory is not well defined. As a consequence, one has to define new measures of synchronization for continuous variable systems like optomechanical arrays~\cite{MarquardtPhysRevLett.111.073603}. These measures are based on quadratures of the coupled systems and allow one to extend the notion of phase synchronization to the quantum regime~\cite{FazioPhysRevLett.111.103605}. Additional measures of synchronization open intriguing connections to concepts of quantum information theory~\cite{LloydRevModPhys.84.621,VanLoockRevModPhys.77.513}, such as decoherence-free subspaces~\cite{Zambrini}, quantum discord~\cite{ZambriniOsc},  entanglement~\cite{LeeWang, ZambriniSpin}, and mutual information~\cite{FazioPhysRevA.91.012301}.
Despite the intensive theoretical investigation of quantum signatures of synchronized states, up to now the quantum manifestations of chimera states are still unresolved.

Recently, we have studied the emergence of chimera states in a network of coupled quantum Van der Pol oscillators \cite{BAS15}, and here we review this work. Unlike in previous studies~\cite{Viennot}, we address the fundamental issue of the dynamical properties of chimera states in a continuous variable system. Considering the chaotic nature of chimera states~\cite{WOL11a}, we study the short-time evolution of the quantum fluctuations at the Gaussian level. This approach allows us to use powerful tools of quantum information theory to describe the correlations in a nonequilibrium state of the system. We show that quantum manifestations of the chimera state appear in the covariance matrix and are related to bosonic squeezing, thus bringing these signatures into the realm of observability in trapped ions~\cite{SadeghpourPhysRevLett.111.234101}, optomechanical arrays~\cite{MarquardtPhysRevLett.111.073603}, and driven-dissipative Bose-Einstein condensates~\cite{DiehlPhysRevLett.110.195301,DiehlPhysRevX.4.021010}. We find that the chimera states can be characterized in terms of R{\'e}nyi  quantum mutual 
information~\cite{AdessoPhysRevLett.109.190502}. Our results reveal that the mutual information for a chimera state lies between the values for synchronized and desynchronized states, which extends in a natural way the definition of chimera states to quantum mechanics.

\section{Nonlinear quantum oscillators}
\label{SecPreliminary}
In this section we describe a recent theoretical proposal to realize a quantum analogue of the Van der Pol oscillator~\cite{SadeghpourPhysRevLett.111.234101, BruderPhysRevLett.112.094102}.
\subsection{The classical Van der Pol oscillator}
\label{ClassicalSingleQuantumVdP}
%
\index{Harmonic oscillator, damped classical}
\index{Van der Pol oscillator, classical}
%
The  classical Van der Pol oscillator is given by the equation of motion~\cite{van1920theory}
\begin{equation}
      \label{ClassSingVdP}
            \ddot{Q}+\omega^2_{0}Q-\epsilon(1-2Q^2)\dot{Q}=0
      \ .
\end{equation}
where $Q \in \mathbb{R}$ is the dynamical variable, $\omega_0$ is the linear frequency, and $\epsilon>0$ is the nonlinearity parameter.
One important aspect of this equation is that the interplay between negative damping proportional to $-\dot{Q}$ and nonlinear damping $Q^2\dot{Q}$ leads to the existence of self-sustained limit cycle oscillations.
Similarly to the method discussed in Ref.~\cite{SadeghpourPhysRevLett.111.234101}, we consider a transformation into a rotating frame $Q(t)=2^{-1/2}(\alpha(t)e^{\mathrm{i}\omega_{0}t}+\alpha^{\ast}(t)e^{-\mathrm{i}\omega_{0}t})$ with a slowly varying complex amplitude $\alpha=2^{-1/2}(Q+\mathrm{i}P)$, and $Q$ and $P=\dot{Q}$ 
denote position and conjugate momentum, respectively. In the rotating frame, one can neglect fast oscillating terms in Eq.~(\ref{ClassSingVdP}) as long as the condition $\epsilon\ll1$ holds. This enables us to obtain an effective amplitude equation which has the form of a Stuart-Landau equation
\begin{equation}
\dot{\alpha}(t)=\frac{\epsilon}{2}(1-|\alpha(t)|^2)\alpha(t)
\end{equation} 
describing the dynamics of the oscillator. In the stationary state, i.e., when $\dot{\alpha}(t)=0$, it admits the existence of a limit cycle defined by $|\alpha(t)|^2=1$.

\subsection{The quantum Van der Pol oscillator}
\label{QuantumSingleQuantumVdP}
%
\index{Harmonic oscillator, damped quantum}
\index{Van der Pol oscillator, quantum}
\index{Master equation}
%
To obtain a quantum analogue of the Van der Pol oscillator, we require a mechanism to inject energy into the system in a linear way (negative damping) and to induce nonlinear losses. Such features can be accomplished by using trapped ions setups~\cite{WinelandRevModPhys.75.281}, as proposed in Ref.~\cite{SadeghpourPhysRevLett.111.234101}. In the case of trapped ions, the dynamic degrees of freedom are described by means of bosonic creation and annihilation operators $a^{\dagger}$ and $a$, respectively. The dissipative dynamics is governed by the Lindblad master equation
\index{Master equation}
%
\begin{equation}
      \label{SingleVDPME}
            \dot{\rho}(t)=2\kappa_{1}\left(a^{\dagger}\rho a-\frac{1}{2}\{\rho,a a^{\dagger}\}\right)+2\kappa_{2}\left(a^{2}\rho (a^{\dagger})^{2}-\frac{1}{2}\{\rho,(a^{\dagger})^{2}a^{2}\}\right)
      \ ,
\end{equation}
in a rotating frame with frequency $\omega_0$, where $\rho$ is the density matrix, and $\kappa_{1}$, $\kappa_{2}$ are dissipation rates.

In the high-photon density limit $\langle a^{\dagger} a\rangle=|\alpha|^{2}\gg 1$ one can describe bosonic quantum fluctuations $\tilde{a},\tilde{a}^{\dagger}$ about the mean field $\alpha$. This approach enables us to study the time evolution of the quantum fluctuations, which is influenced by the mean field solution. Correspondingly, at mean-field level, the system resembles the classical behavior of the Van der Pol oscillator in the $\epsilon\ll1$ limit. Unfortunately, to obtain analytical results we must confine ourselves to the study of Gaussian quantum fluctuations. This implies some limitations in the description of the long-time dynamics of the fluctuations. For example, even if one prepares the system in a coherent state at $t=0$ $\rho(0)=|\alpha(0)\rangle\langle\alpha(0)|$, i.e., a bosonic Gaussian state, there are quantum signatures of the classical limit cycle leading to non-Gaussian effects. Therefore, within the framework of a Gaussian description, one is able to describe only short-time 
dynamics, where the non-Gaussian effects are negligible.

\subsection{Gaussian quantum fluctuations and semiclassical trajectories}
\label{QuantumFluctuationVdP}
\index{Quantum fluctuations}
\index{Semiclassical dynamics}
Let us begin by considering the decomposition $a(t)=\tilde{a}+\alpha(t)$ of the bosonic operator $a$ in terms of the quantum fluctuations $\tilde{a}$ and the mean field $\alpha$.
For completeness, in appendix~\ref{appendix} we calculate explicitly the Gaussian quantum fluctuations for the quartic oscillator.
To formalize this procedure from the perspective of the master equation~\cite{ArmouPhysRevLett.104.053601,HammererPhysRevX.4.011015}, we define the density matrix in the co-moving frame $\rho_{\alpha}(t)=\hat{D}^{\dagger}\left[\alpha(t)\right]\rho(t)\hat{D}\left[\alpha(t)\right]$, where $\hat{D}\left[\alpha(t)\right]=\exp\left[\alpha(t)  \tilde{a}^{\dagger}-\alpha^{*}(t)\tilde{a}\right]$ is a displacement operator.
In the co-moving frame, we obtain a master equation with Liouville operators $\hat{\mathcal{L}}_{1}$ and $\hat{\mathcal{L}}_{2}$
\index{Master equation}
\index{Liouville operator}
%
\begin{align}
      \label{VDPMasterEqRot}
        \dot{\rho}_{\alpha}(t)&=-\mathrm{i}[\hat{H}^{(\alpha)}(t),\rho_{\alpha}(t)]+2\kappa_{1}\hat{D}^{\dagger}\left[\alpha(t)\right]\left(\tilde{a}^{\dagger}\rho \tilde{a}-\frac{1}{2}\{\rho,\tilde{a} \tilde{a}^{\dagger}\}\right)\hat{D}\left[\alpha(t)\right] 
        \nonumber \\&
        +2\kappa_{2}\hat{D}^{\dagger}\left[\alpha(t)\right]\left(\tilde{a}^{2}\rho (\tilde{a}^{\dagger})^{2}-\frac{1}{2}\{\rho,(\tilde{a}^{\dagger})^{2}\tilde{a}^{2}\}\right)\hat{D}\left[\alpha(t)\right] 
        \nonumber \\&
        \equiv-\mathrm{i}[\hat{H}^{(\alpha)}(t),\rho_{\alpha}(t)]+\hat{\mathcal{L}}_{1}\rho_{\alpha}+\hat{\mathcal{L}}_{2}\rho_{\alpha}
      \ ,
\end{align}
where we have defined $\hat{H}^{(\alpha)}(t)=-\mathrm{i}\hat{D}^{\dagger}\left[\alpha(t)\right]\partial_t\hat{D}\left[\alpha(t)\right]$ and the anticommutator  $\{\hat{A},\hat{B}\}=\hat{A}\hat{B}+\hat{B}\hat{A}$. In the co-moving frame, the coherent dynamics is generated by the Hamiltonian
\begin{equation}
      \label{DerDisp}
          \hat{H}^{(\alpha)}(t)
          =-\frac{\mathrm{i}}{2}[\dot{\alpha}(t)\alpha^{*}(t)-\alpha(t)\dot{\alpha}^{*}(t)]-\mathrm{i}[\dot{\alpha}(t)\tilde{a}^{\dagger}-\dot{\alpha}^{*}(t)\tilde{a}]
      \ .
\end{equation}

A next step in the calculation of the Gaussian quantum fluctuations is to expand the Liouville operators $\hat{\mathcal{L}}_{1}$ and $ \hat{\mathcal{L}}_{2}$ in terms of the quantum fluctuations. We begin by considering the Liouville operator $\hat{\mathcal{L}}_{1}$, which preserves the Gaussian character of the initial state $\rho(0)=|\alpha(0)\rangle\langle\alpha(0)|$.
By using elementary properties of the displacement operator we can decompose the dissipative term into coherent and incoherent parts:
\index{Liouville operator}
%
\begin{align}
      \label{DisplLinDiss}
           \hat{\mathcal{L}}_{1}\rho_{\alpha}(t)&=2\kappa_{1}\hat{D}^{\dagger}\left[\alpha(t)\right]\left(\tilde{a}^{\dagger}\rho \tilde{a}-\frac{1}{2}\{\rho,\tilde{a} \tilde{a}^{\dagger}\}\right)\hat{D}\left[\alpha(t)\right]
           \nonumber\\
           &=2\kappa_{1}\left(\tilde{a}^{\dagger}\rho_{\alpha} \tilde{a}-\frac{1}{2}\{\rho_{\alpha},\tilde{a} \tilde{a}^{\dagger}\}\right)-\mathrm{i}\left[\mathrm{i}\kappa_{1}\alpha \tilde{a}^{\dagger},\rho_{\alpha}\right]-\mathrm{i}\left[-\mathrm{i}\kappa_{1}\alpha^{*}\tilde{a},\rho_{\alpha}\right]
      \ .
\end{align}
In the calculation of the dissipative term proportional to $\kappa_{2}$, one needs to be particularly careful, because it causes non-Gaussian effects due to two-photon processes. Interestingly, one can decompose the Liouville operator $\hat{\mathcal{L}}_{2}$ into coherent terms given by commutators of an effective Hamiltonian with the density operator and terms preserving the Gaussian character of the initial state. In addition, we also obtain explicitly the non-Gaussian contributions:
\index{Liouville operator}
%
\begin{align}
      \label{DisplNonLinDiss}
           \hat{\mathcal{L}}_{2}\rho_{\alpha}(t)&=2\kappa_{2}\hat{D}^{\dagger}\left[\alpha(t)\right]\left(\tilde{a}^{2}\rho (\tilde{a}^{\dagger})^{2}-\frac{1}{2}\{\rho,(\tilde{a}^{\dagger})^{2}\tilde{a}^{2}\}\right)\hat{D}\left[\alpha(t)\right]
           \nonumber\\
           &=2\kappa_{2}\left(\tilde{a}^{2}\rho_{\alpha} (\tilde{a}^{\dagger})^{2}-\frac{1}{2}\{\rho_{\alpha},(\tilde{a}^{\dagger})^{2}\tilde{a}^{2}\} \right)+4\kappa_{2}\alpha^{*}\left(\tilde{a}^{2}\rho_{\alpha} \tilde{a}^{\dagger}-\frac{1}{2}\{\rho_{\alpha},\tilde{a}^{\dagger}\tilde{a}^{2}\} \right)
           \nonumber\\&
           +4\kappa_{2}\alpha\left(\tilde{a}\rho_{\alpha} (\tilde{a}^{\dagger})^{2}-\frac{1}{2}\{\rho_{\alpha},(\tilde{a}^{\dagger})^{2}\tilde{a}\} \right)
           +8\kappa_{2}|\alpha|^2\left(\tilde{a}\rho_{\alpha}\tilde{a}^{\dagger}-\frac{1}{2}\{\rho_{\alpha},\tilde{a}^{\dagger}\tilde{a}\} \right)
           \nonumber\\&
           -\mathrm{i}\left[\mathrm{i}\kappa_{2}(\alpha^{*})^{2}\tilde{a}^{2},\rho_{\alpha}\right]-\mathrm{i}\left[-\mathrm{i}\kappa_{2}\alpha^{2}(\tilde{a}^{\dagger})^{2},\rho_{\alpha}\right]
           \nonumber\\&
           -\mathrm{i}\left[2\mathrm{i}\kappa_{2}\alpha (\alpha^{*})^2 \tilde{a},\rho_{\alpha}\right]-\mathrm{i}\left[-2\mathrm{i}\kappa_{2}\alpha^{*}\alpha^2\tilde{a}^{\dagger},\rho_{\alpha}\right]
      \ .
\end{align}
In the semiclassical high-density limit $|\alpha|^2\gg 1$, one can safely neglect the effect of the non-Gaussian terms in Eqs.~\eqref{DisplLinDiss} and~\eqref{DisplNonLinDiss}. Furthermore, we require vanishing linear terms in the coherent part of the master equation~\eqref{VDPMasterEqRot}. This is achieved as long as the condition
\index{Van der Pol oscillator, classical}
%

\begin{equation}
      \label{QuantEqMotion}
            \dot{\alpha}(t)=\kappa_{1}\alpha(t)-2\kappa_{2}\alpha(t)|\alpha(t)|^2
\end{equation}
is satisfied. After neglecting such terms, we obtain the master equation
\begin{align}
      \label{EffVDPMasterEqRot}
        \dot{\rho}_{\alpha}(t)&=-\mathrm{i}\kappa_{2}\left[\mathrm{i}(\alpha^{*})^{2}\tilde{a}^{2}-\mathrm{i}\alpha^{2}(\tilde{a}^{\dagger})^{2},\rho_{\alpha}\right]+2\kappa_{1}\left(\tilde{a}^{\dagger}\rho \tilde{a}-\frac{1}{2}\{\rho,\tilde{a} \tilde{a}^{\dagger}\}\right)
        \nonumber \\& 
        +8\kappa_{2}|\alpha(t)|^2\left(\tilde{a}\rho_{\alpha}\tilde{a}^{\dagger}-\frac{1}{2}\{\rho_{\alpha},\tilde{a}^{\dagger}\tilde{a}\} \right)
      \ .
\end{align}
One can interpret this procedure from a geometrical point of view. An initial coherent state $\rho(0)=|\alpha(0)\rangle\langle\alpha(0)|$ corresponds to the vacuum $\rho_{\alpha}(0)=|0\rangle\langle 0|$ in the co-moving frame. On the other hand, $\alpha(0)$ plays the role of the initial condition for the classical equation of motion Eq.~\eqref{QuantEqMotion}. The solution $\alpha(t)$ of Eq.~\eqref{QuantEqMotion} is responsible for the emergence of time dependent rates in the master equation and time dependent squeezing. In the stationary limit, however, the classical equations of motion exhibit self-sustained oscillations with amplitude $|\alpha(t)|^2=\kappa_{1}/2\kappa_{2}$. In this limit, the master equation ceases to have time-dependent coefficients. 

\section{Quantum description of a network of coupled Van der Pol oscillators}
\label{QuantumChimeraVdP}
We consider a quantum network consisting of a ring of $N$ coupled Van der Pol oscillators.
Such a network can be described by the master equation for the density matrix $\rho(t)$
\index{Master equation}
\index{Liouville operator}
%
\begin{align}
      \label{MasterEqNetwork}
            \dot{\rho}&=-\frac{\mathrm{i}}{\hbar}[\hat{H},\rho]+2\sum^{N}_{l=1}\left[\kappa_{1}\mathcal{D}(a_{l}^{\dagger})+\kappa_{2}\mathcal{D}(a_{l}^{2})\right]
      \ ,
\end{align}
where $a^{\dagger}_{l},a_{l}$ are creation and annihilation operators of bosonic particles and  $\mathcal{D}(\hat{O})=\hat{O}\rho\hat{O}^{\dagger}-\frac{1}{2}(\hat{O}^{\dagger}
\hat{O}\rho+\rho\hat{O}^{\dagger}
\hat{O})$ describes dissipative processes with rates $\kappa_1,\kappa_{2}>0$.
In addition, we have imposed periodic boundary conditions $a_{l}=a_{l+N}$ for the bosonic operators.
In contrast to Ref.~\cite{SadeghpourPhysRevLett.111.234101}, we consider a nonlocal coupling between the oscillators.
Therefore, the Hamiltonian in the interaction picture reads $\hat{H}=\hbar\sum^{N}_{l\neq m=1}K_{l,m}(a^{\dagger}_{l}a_{m}+a_{l}a^{\dagger}_{m})$,
where $K_{l,m}=\frac{V}{2d}\Theta(d-|l-m|)$ is the coupling matrix of the network and $\Theta(x)$ is the Heaviside step function. This kind of coupling implies that  Eq.~\eqref{MasterEqNetwork} has a rotational $S^1$ symmetry as discussed in Ref.~\cite{ZAK14}. One can include counter-rotating terms such as $a^{\dagger}_{l}a^{\dagger}_{m}$ in the coupling, but this would lead to symmetry breaking.

This definition implies that the coupling is zero if the distance $|l-m|$  between the $l$-th and $m$-th the nodes is bigger than the coupling range $d$. On the other hand, if $|l-m|<d$,  then $K_{l,m}=\frac{V}{2d}$. 
In the particular case $d=N/2$ one has all-to-all coupling and recovers the results of Ref.~\cite{SadeghpourPhysRevLett.111.234101}.

Now we compare Eq.\eqref{MasterEqNetwork} with the general form of the Lindblad master equation ~\cite{carmichael2009statistical}
\begin{equation}
      \label{LimbladVDP}
            \dot{\rho}(t)=-\frac{\mathrm{i}}{\hbar}[\hat{H},\rho]+\sum_{\mu}\gamma_{\mu}\left(\hat{L}_{\mu}\rho \hat{L}_{\mu}^{\dagger}-\frac{1}{2}\{\rho,\hat{L}_{\mu}^{\dagger}\hat{L}_{\mu}\}\right)
\end{equation}
with Lindblad operators $\hat{L}_{\mu}$. This enables us to introduce an effective Hamiltonian which describes the dynamics between quantum jumps
\begin{equation}
      \label{EffGenHam}
            \hat{H}_{\text{eff}}=\hat{H}-\frac{\mathrm{i}\hbar}{2}\sum_{\mu}\gamma_{\mu}\hat{L}_{\mu}^{\dagger}\hat{L}_{\mu}
      \ .
\end{equation}
In the case of the master equation Eq.\eqref{MasterEqNetwork}, the effective Hamiltonian reads
\begin{align}
      \label{NonHermBoseHubbard}
            \hat{H}_{\text{eff}}&=-\mathrm{i}\hbar\kappa_{1}\sum_{l=1}^{N}(a^{\dagger}_{l}a_{l}+1)-\mathrm{i}\kappa_{2}\sum_{l=1}^{N}\hat{n}_{l}(\hat{n}_{l}-1)
            \nonumber \\&
            +\hbar\sum^{N}_{l\neq m=1}K_{l,m}(a^{\dagger}_{l}a_{m}+a_{l}a^{\dagger}_{m})
      ,
\end{align}
where $\hat{n}_{l}=a^{\dagger}_{l}a_{l}$. The Hamiltonian Eq.~\eqref{NonHermBoseHubbard} describes a  Bose-Hubbard model with long range interactions, where on-site energies and chemical potential are complex. This kind of model arises naturally in the context of driven-dissipative Bose-Einstein condensation~\cite{DiehlPhysRevLett.110.195301,DiehlPhysRevX.4.021010}.

\subsection{Gaussian quantum fluctuations and master equation}
\label{GuaussianQuantumFlucChimeraVdP}
%
\index{Quantum fluctuations}

We define the expansion $b_{l}(t)= \hat{D}^{\dagger}\left[\bm{\alpha}(t)\right]a_{l}\hat{D}\left[\bm{\alpha}(t)\right]=\tilde{a}_{l}+\alpha_{l}(t)$, where 
$\hat{D}\left[\bm{\alpha}(t)\right]=\exp\left[\bm{\alpha}(t) \cdot \hat{\bm{\tilde{a}}}^{\dagger}-\bm{\alpha}^{*}(t)\cdot\hat{\bm{\tilde{a}}}\right]$, $\bm{\alpha}(t)=[\alpha_1(t),\ldots,\alpha_{N}(t)]$ and $\hat{\bm{\tilde{a}}}=(\tilde{a}_{1},\ldots,\tilde{a}_{N})$ as in Ref.~\cite{GlauberPhysRev.177.1857}.
In this work we consider the semiclassical regime, where the magnitude of the mean field $\alpha_{l}(t)$ is larger than the quantum fluctuations $\tilde{a}_{l}$ as in Refs.~\cite{ArmouPhysRevLett.104.053601,HammererPhysRevX.4.011015}.  
By using the expansion of the master equation about the mean-field  $\bm{\alpha}(t)$ described in the previous section, we obtain a master equation for the  density operator in a co-moving frame $\rho_{\bm{\alpha}}(t)=\hat{D}^{\dagger}\left[\bm{\alpha}(t)\right]\rho(t)\hat{D}\left[\bm{\alpha}(t)\right]$ 
\index{Master equation}
\index{Liouville operator}
%
\begin{equation}
      \label{SemiclassMasterNetwork}
        \dot{\rho}_{\bm{\alpha}}\approx-\frac{\mathrm{i}}{\hbar}[\hat{H}_{\text{Q}}^{(\bm{\alpha})},\rho_{\bm{\alpha}}]+2\sum^{N}_{l=1}\left[\kappa_{1}\mathcal{D}(\tilde{a}_{l}^{\dagger})
        +4\kappa_{2}|\alpha_{l}|^2\mathcal{D}(\tilde{a}_{l})\right]
      \ .
\end{equation}
In addition, the coherent dynamics of the fluctuations are governed by
the Hamiltonian
\begin{align}
      \label{EffQuadHamNet}
            \hat{H}^{(\bm{\alpha})}_{\text{Q}}(t)&=\sum^{N}_{l=1}\left(\mathrm{i}\kappa_{2}(\alpha_{l}^{*})^{2}\tilde{a}_{l}^{2}-\mathrm{i}\kappa_{2}\alpha_{l}^{2}(\tilde{a}_{l}^{\dagger})^{2}+\frac{1}{N}\sum^{N}_{r=1}K_{l,l+r}(\tilde{a}^{\dagger}_{l}\tilde{a}_{l+r}+\tilde{a}_{l}\tilde{a}^{\dagger}_{l+r})\right)
            \nonumber\\&
            +\sum^{N}_{l=1}\left(-\mathrm{i}[\dot{\alpha}_{l}(t)\tilde{a}_{l}^{\dagger}-\dot{\alpha}_{l}^{*}(t)\tilde{a}_{l}]+\mathrm{i}\kappa_{1}\alpha_{l} \tilde{a}_{l}^{\dagger}
            -\mathrm{i}\kappa_{1}\alpha_{l}^{*}\tilde{a}\right)
            \nonumber\\&
            +\sum^{N}_{l=1}\left(
            2\mathrm{i}\kappa_{2}\alpha_{l} (\alpha_{l}^{*})^2 \tilde{a}_{l}-2\mathrm{i}\kappa_{2}\alpha_{l}^{*}\alpha_{l}^2\tilde{a}_{l}^{\dagger}\right)
            \nonumber\\&
            +\sum^{N}_{r=1}K_{l,l+r}(\tilde{a}^{\dagger}_{l}\alpha_{l+r}+\tilde{a}_{l}\alpha^{*}_{l+r}+\tilde{a}_{l+r}\alpha^{*}_{l}+\tilde{a}^{\dagger}_{l+r}\alpha_{l})
      \ .
\end{align}
To obtain the equations for the mean field, the linear terms in the expansion Eq.~\eqref{EffQuadHamNet} must vanish, which leads to the equation
\index{Chimera states, quantum}
\index{Chimera states, classical}
%
\begin{equation}
      \label{QuantEqMotionNetwork}
            \dot{\alpha}_{l}(t)=\alpha_{l}(t)(\kappa_{1}-2\kappa_{2}|\alpha_{l}(t)|^2)-\mathrm{i}\sum^{N}_{s\neq l}K_{l,s}\alpha_{s}(t)
\end{equation}
with a similar equation for $\dot{\alpha}^{*}_{l}(t)$. Finally, by using the master equation \eqref{SemiclassMasterNetwork} we can calculate the equations of motion as follows
\begin{align}
      \label{ExpValGaussianFluc}
             \frac{d \langle \tilde{a}_i\rangle}{d t}&=\text{tr}[\tilde{a}_{i}\dot{\rho}_{\bm{\alpha}}(t)]=\kappa_1\langle \tilde{a}_i\rangle-4\kappa_2|\alpha_{i}|^2\langle \tilde{a}_i\rangle-2\kappa_2\alpha^{2}_{i}\langle \tilde{a}^{\dagger}_i\rangle-\mathrm{i}\sum^{N}_{s\neq i}K_{i,s}\langle \tilde{a}_s\rangle
      \ .
\end{align}

\subsection{Relation to the continuum limit and linearization}
\label{ContinuumLimitQuantumChimeraVdP}
%
\index{Chimera states, classical}
To understand the meaning of the Gaussian quantum fluctuations we discuss the continuum limit of the classical equations of motion Eq.~\eqref{QuantEqMotionNetwork}. 
In the continuum limit, $N\rightarrow\infty$, the complex variable $\alpha_{l}(t)=r_{l}(t)e^{\mathrm{i}\phi_{l}(t)}$ can be described by means of a complex field $\alpha(x,t)=|\alpha(x,t)|e^{\mathrm{i}\phi(x,t)}$, where $x$ is the continuous version of the index $l$. Correspondingly, $|\alpha(x,t)|$ and $\phi(x,t)$ represent the amplitude and phase fields, respectively~\cite{KUR02a}.

In the continuum limit, a ring of $N$ coupled nodes can be described by means of a classical field $\alpha(x,t)$ defined on a circle of length $L$, where $x$ is the position coordinate. In addition, if one introduces the continuum version $K(x-y)$ of the coupling matrix $K_{l,m}$, the dynamics of such a field can be described by the equation of motion 
\index{Phase oscillators}
%
\begin{equation}
      \label{EqMotClassField}
            \frac{\partial \alpha(x,t)}{\partial t}=\alpha(x,t)(\kappa_{1}-2\kappa_{2}|\alpha(x,t)|^2)-\mathrm{i}\int^{L}_{0}dy\ K(x-y)\alpha(y,t)
      \ ,
\end{equation}
which  is the continuum limit of Eq.~\eqref{QuantEqMotionNetwork}  and resembles the field equation discussed in Ref.~\cite{KUR02a}. In particular, if we assume the amplitude $|\alpha(x,t)|=r_{0}$ to be constant after the system is trapped into the limit cycle, we obtain a differential equation for the phases
\begin{equation}
      \label{EqMotPhasesClassField}
            \mathrm{i}\frac{\partial \phi(x,t)}{\partial t}=(\kappa_{1}-2\kappa_{2}r^{2}_{0})-\mathrm{i}\int^{L}_{0}dy\ K(x-y)e^{-\mathrm{i}[\phi(x,t)-\phi(y,t)]}
      \ ,
\end{equation}
Following the method described in Ref.~\cite{KUR02a}, one can introduce a mean field
\begin{equation}
      \label{ClassKuramotoOrderPar}
            r(x,t)e^{\mathrm{i}\Theta(x,t)}=\int^{L}_{0}dy\ K(x-y)e^{\mathrm{i}\phi(y,t)}
\end{equation}
This method works well in the case of phase chimeras. However, in the case of amplitude-mediated chimeras~\cite{SET13,SET14}, one requires to study both phase $\phi(x,t)$ and amplitude $|\alpha(x,t)|$ fields.

In order to have a better understanding of the quantum fluctuations, let us linearize the equation of motion for the field Eq.~\eqref{EqMotClassField} about a solution $\alpha_{0}(x,t)$. For this purpose, let us consider the decomposition of the field $\alpha(x,t)=\alpha_{0}(x,t)+\tilde{a}(x,t)$, where $\tilde{a}(x,t)$ is a small perturbation such that $|\alpha_{0}(x,t)|\ll |\tilde{a}(x,t)| $. Now let us assume that we expand Eq.~\eqref{EqMotClassField} up to first order in the perturbation. After some algebraic manipulations we obtain
\begin{align}
      \label{LinPerAnaly}
            \frac{\partial \tilde{a}(x,t)}{\partial t}&=
            \kappa_{1}\tilde{a}(x,t)-4\kappa_{2}|\alpha_{0}(x,t)|^{2}\tilde{a}(x,t)
            \nonumber\\&
            -2\kappa_{2}[\alpha_0(x,t)]^2\tilde{a}^{*}(x,t)-\mathrm{i}\int^{L}_{0}dy\ K(x-y)\tilde{a}(y,t)
      \ .
\end{align}
One can observe that this equation is precisely the continuum limit of the equations of motions Eq.~\eqref{ExpValGaussianFluc} for the expectation values of the quantum fluctuations. In particular, $\alpha_0(x,t)$ plays the role of the mean field $\alpha_{l}(t)$ and $\tilde{a}(x,t)$ is the continuum limit of the expectation value $\langle \tilde{a}_l(t)\rangle$

Now the role of the Gaussian quantum fluctuations is clear: By neglecting non-Gaussian contributions in the master equation~\cite{ArmouPhysRevLett.104.053601,HammererPhysRevX.4.011015}, one constructs the master equation~\eqref{SemiclassMasterNetwork} governing the evolution of the quantum fluctuations. Due to the chaotic nature of the classical chimera states~\cite{WOL11a}, one expects giant quantum fluctuations about the semiclassical trajectories~\cite{HaakePhysRevLett.108.073601}. As a consequence, the Gaussian approximation, i.e., the master equation Eq.~\eqref{SemiclassMasterNetwork} fails to describe the long-time dynamics.

\subsection{The Gutzwiller ansatz and the master equation}
\label{GutzwillerQuantumChimeraVdP}
%
\index{Quantum fluctuations}
\index{Gutzwiller ansatz}
In this section we describe the different methods to tackle the emergence of chimera states in the quantum regime. Due to the nature of the nodes of the network, one has to truncate the Hilbert space up to a certain occupation number $n_t$ of the oscillator. This implies that if one has a network with $N$ nodes, one has to solve a system of $n_{t}^{2N}$ coupled differential equations for the elements $\rho_{n,m}$ of the density matrix as follows from Eq.~\eqref{MasterEqNetwork}. Chimera states usually emerge in networks consisting at least of $N=40$ nodes. This means that if one truncates the Hilbert space of the oscillator up to $n_t=2$ one has to solve a system of $3^{80}$ coupled differential equations. In the incoherent regime of the network one expects vanishing coherences. As a consequence of this one has to solve only the evolution of the populations, which involves the solution of $3^{40}$ ordinary differential equations. 
From the previous analysis we conclude that the complete solution of the master equation \eqref{MasterEqNetwork} is not possible in order to find the quantum signatures of synchronization and even to describe the incoherent regime. Therefore one has to invoke alternative methods of solution as we describe below.

We start by considering a mean-field ansatz of the density matrix $\rho(t)=\bigotimes_{l=1}^{N}\rho_{l}(t)$. Similarly to  Refs.~\cite{MarquardtPhysRevLett.111.073603,SadeghpourPhysRevLett.111.234101}, we obtain a self-consistent system of equations
\index{Master equation}
\index{Liouville operator}
%
\begin{equation}
      \label{SelfConsMaster}
            \dot{\rho}_{l}(t)=-\frac{\mathrm{i}}{\hbar}[\hat{H}_{l},\rho_{l}]+2\kappa_{1}\mathcal{D}(a_{l}^{\dagger})+\kappa_{2}\mathcal{D}(a_{l}^{2})
      \ ,
\end{equation}
where we define the self-consistent local Hamiltonian
\begin{equation}
      \label{SelfConsHamiltonian}
            \hat{H}_{l}=\hbar \Gamma_{l} a^{\dagger}_{l}+\hbar \Gamma^{\ast}_{l} a_{l}.
\end{equation}
%
\index{Chimera states, quantum}
Motivated by the original approach of Kuramoto~\cite{KUR02a} and a recent work~\cite{LeeWang}, we have defined the mean field $\Gamma_{l}=\hbar\sum^{N}_{r=1}K_{l,l+r}\langle a_{l+r}\rangle$, which resembles the order parameter Eq.~\eqref{ClassKuramotoOrderPar}. This order parameter takes into account the contributions of the quantum coherences.

For completeness, we discuss briefly the semiclassical equations of motion derived from Eq.~\eqref{MasterEqNetwork}
\begin{equation}
      \label{SemiclassEq}
            \frac{d \langle a_l\rangle}{d t}=\text{tr}[a_{l}\dot{\rho}(t)]=\kappa_{1}\langle a_l\rangle-2\kappa_{2}\langle a^{\dagger}_l a^{2}_l\rangle-\frac{\mathrm{i}}{N}\sum^{N}_{r=1}K_{l,l+r}\langle a_{l+r}\rangle
       \ .
\end{equation}
Interestingly, the equations of motion Eq.~\eqref{ExpValGaussianFluc} constitute a particular case of Eq.~\eqref{SemiclassEq}, because if one only considers the contributions of the Gaussian fluctuations, one obtains a natural way to factorize expectation values~\cite{MarquardtPhysRevLett.111.073603,SadeghpourPhysRevLett.111.234101}.
In contrast to the complete solution of Eq.~\eqref{MasterEqNetwork}, the Gutzwiller ansatz allows one to obtain  the solution of the problem with polynomial resources. More specifically, instead of solving $3^{2N}$ equations, one has to solve $3^{2}N$ equations if one truncates the bosonic Hilbert space at $n_t=2$. The minimal size of a chain that supports chimeras is of the order of $N=40$, therefore one has to solve only $360$ equations of motion.
\section{Classical chimera states and phase-space methods}
\label{ClassChimPhSpMethNetworkVdP}
%
\index{Chimera states, classical}
The equations of motion Eq.~\eqref{QuantEqMotionNetwork} resemble a system of coupled Stuart-Landau oscillators~\cite{ZAK14}.
The solution $\bm{\alpha}(t)$ of the equations of motion Eq.~\eqref{QuantEqMotionNetwork}, provide us information about the dynamics of the mean field. Such a mean field plays a fundamental role in the study of the master equation  Eq.~\eqref{SemiclassMasterNetwork}. Within the Gaussian approximation, the mean field is responsible for coherent effects such as squeezing in Eq.~\eqref{EffQuadHamNet}. In addition,  the amplitude $|\alpha_{l}(t)|$  determines the dissipation rates which appear in Eq.~\eqref{SemiclassMasterNetwork}. Therefore to investigate the evolution of the density matrix, we require the time evolution  $\bm{\alpha}(t)$. In this section we show that the mean field exhibits chimera-like dynamics. By using phase-space methods, we investigate quantum signatures of these states.
\subsection{Emergence of classical chimera states}
\label{ClassChimNetworkVdP}
\begin{figure}
\centering
\includegraphics[width=0.99\textwidth,clip=true]{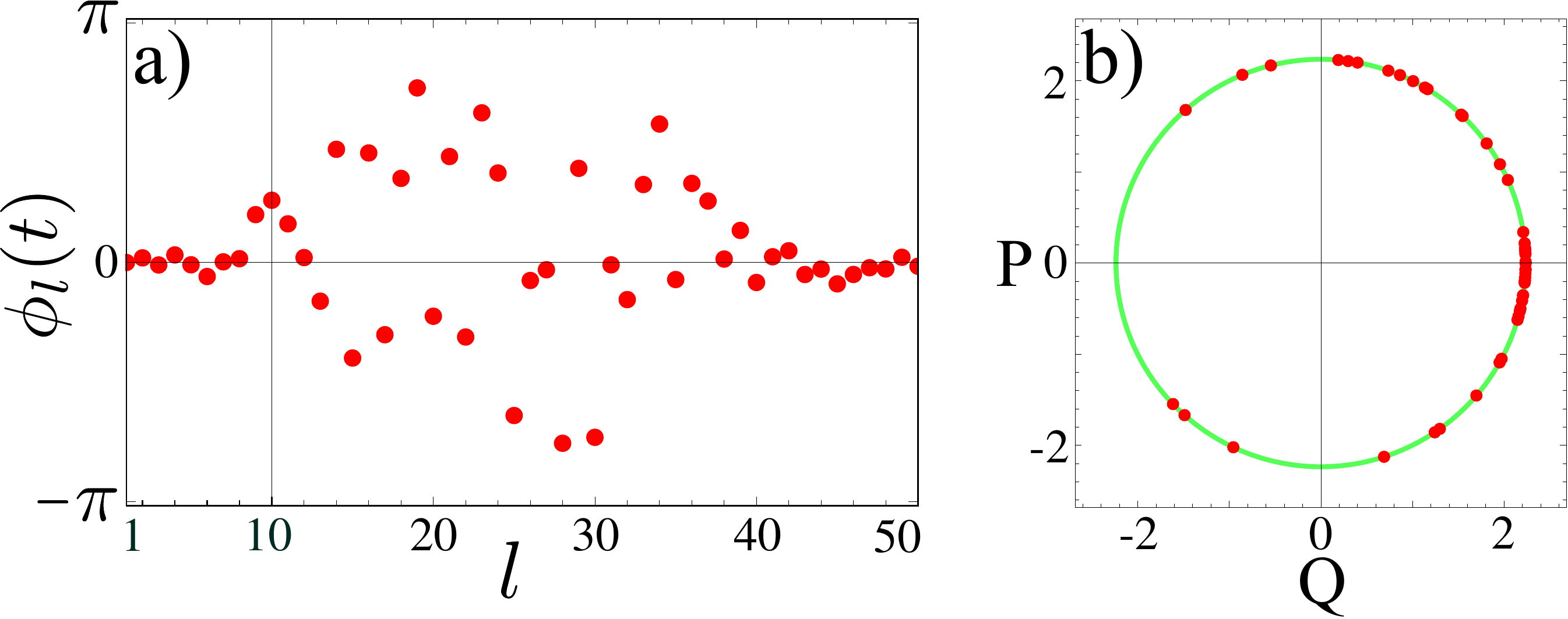}
    \caption{
    \label{FigI1}
   Initial conditions used in the simulations. a) Initial distribution of the phases $\phi_{l}(0)$ drawn randomly from a Gaussian distribution in space.
   b) Phase-space representation of the initial conditions for the oscillators. The green circle represents the limit cycle with radius $|\alpha_{l}(0)|\approx1.58$, where $\alpha_{l}(t)=\frac{Q_{l}(t)+\mathrm{i}P_{l}(t)}{\sqrt{2\hbar}}$. 
   Parameters $\hbar=1$, $d=10$, $\kappa_2=0.2\kappa_1$, and $N=50$.}
\end{figure}

%
\index{Chimera states, classical}
In the case of the uncoupled system $V=0$, i.e., $K_{l,m}=0$, a single Van der Pol oscillator~\cite{SadeghpourPhysRevLett.111.234101, BruderPhysRevLett.112.094102} exhibits a limit cycle with radius $r_{0}=\sqrt{\frac{\kappa_{1}}{2\kappa_{2}}}=1.58$ for the parameters $\kappa_{2}=0.2\kappa_{1}$. For convenience, in the coupled system we consider initial conditions at $t=0$ in such a way that each oscillator has the same amplitude $|\alpha_{l}(0)|\approx1.58$.
In addition, we consider phases drawn randomly from a Gaussian distribution in space
$\phi_{l}(0)=\frac{\theta}{\sqrt{2\pi}\sigma}\exp[-\frac{(l-\mu)^2}{2\sigma^2}]$, where $-24\pi<\theta<24\pi$ is a random number, $\mu=N/2$ and $\sigma=9$. Fig.~\ref{FigI1}a) shows the initial conditions. In terms of the coordinates $\alpha_{l}(t)=\frac{Q_{l}(t)+\mathrm{i}P_{l}(t)}{\sqrt{2\hbar}}$, the initial conditions must satisfy $\sqrt{Q^{2}_{l}(0)+P^{2}_{l}(0)}\approx 2.24$, which defines the green circle in Fig.~\ref{FigI1}b). 

Besides the description of chimera states, we also discuss completely synchronized and completely desynchronized solutions. To obtain such solutions, we consider different coupling strengths $V$. However, for every case, we restrict ourselves to the same initial conditions as in Fig.~\ref{FigI1}.
\begin{figure}
\centering
\includegraphics[width=0.99\textwidth,clip=true]{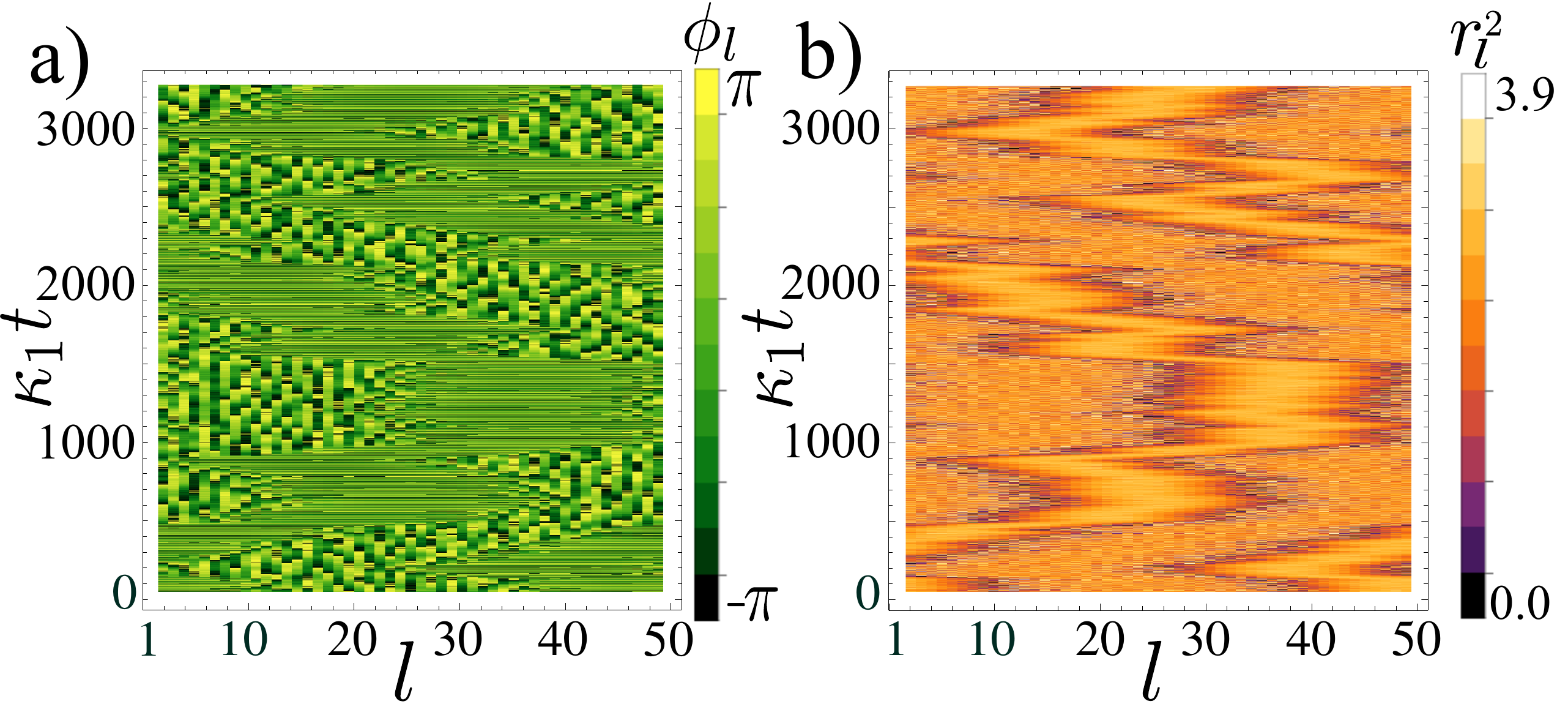}
    \caption{
    \label{Fig1}
   Space-time representation of the classical chimera state. We consider a representation 
   $\alpha_{l}(t)=r_{l}(t)e^{\mathrm{i}\phi_{l}(t)}$ of the individual oscillators in terms of amplitude $r_{l}(t)$ and phase $\phi_{l}(t)$. (a) Depicts the time evolution of the phase chimera, where we represent the phases $\{\phi_{l}(t)\}$. Similarly, (b) depicts the amplitudes $\{r^{2}_{l}(t)\}$.  Parameters: $d=10$, $\kappa_2=0.2 \kappa_1$, $V=1.2\kappa_1$, and $N=50$ \cite{BAS15}.}
\end{figure}

From our previous discussion in section~\ref{GuaussianQuantumFlucChimeraVdP}, the classical equations of motion Eq.~\eqref{QuantEqMotionNetwork} must be satisfied in order to investigate the evolution of the master equation Eq.~\eqref{SemiclassMasterNetwork} in the co-moving frame. In the polar representation $\alpha_{l}(t)=r_{l}(t)e^{\mathrm{i}\phi_{l}(t)}$,
the equations of motion couple amplitude $r_{l}(t)$ and phase $\phi_{l}(t)$ of the individual oscillators. 
We numerically solve Eq.~\eqref{QuantEqMotionNetwork} for a network of $N=50$ coupled oscillators with coupling range $d=10$. We consider initial conditions $|\alpha_{l}(t_{0})|\approx r_{0}$, where $r_{0}=1.58$, and phases drawn randomly from a Gaussian distribution in space. Figure~\ref{Fig1} depicts the time evolution of a classical chimera state. 
In Fig.~\ref{Fig1}~(a) we show the space-time representation of the phases $\phi_{l}(t)$ of the individual oscillators. One can observe that for a fixed time, there is a domain of
synchronized oscillators that coexists with a domain of desynchronized motion, which is a typical feature of chimera states. Besides the phase, also the amplitude exhibits chimera dynamics as we show in Fig.~\ref{Fig1}~(b). 
One can observe that the width of the synchronized region changes with time. Similarly, the center of mass of the synchronized region moves randomly along the ring.  Chimera states with these features have been reported in the literature and are referred to as \textit{breathing} and  \textit{drifting}
chimeras~\cite{WOL11a}.

\subsection{Solution of the Fokker-Planck equation}
\label{FokkerPlanckVdP}
\begin{figure}
\centering
\includegraphics[width=0.99\textwidth,clip=true]{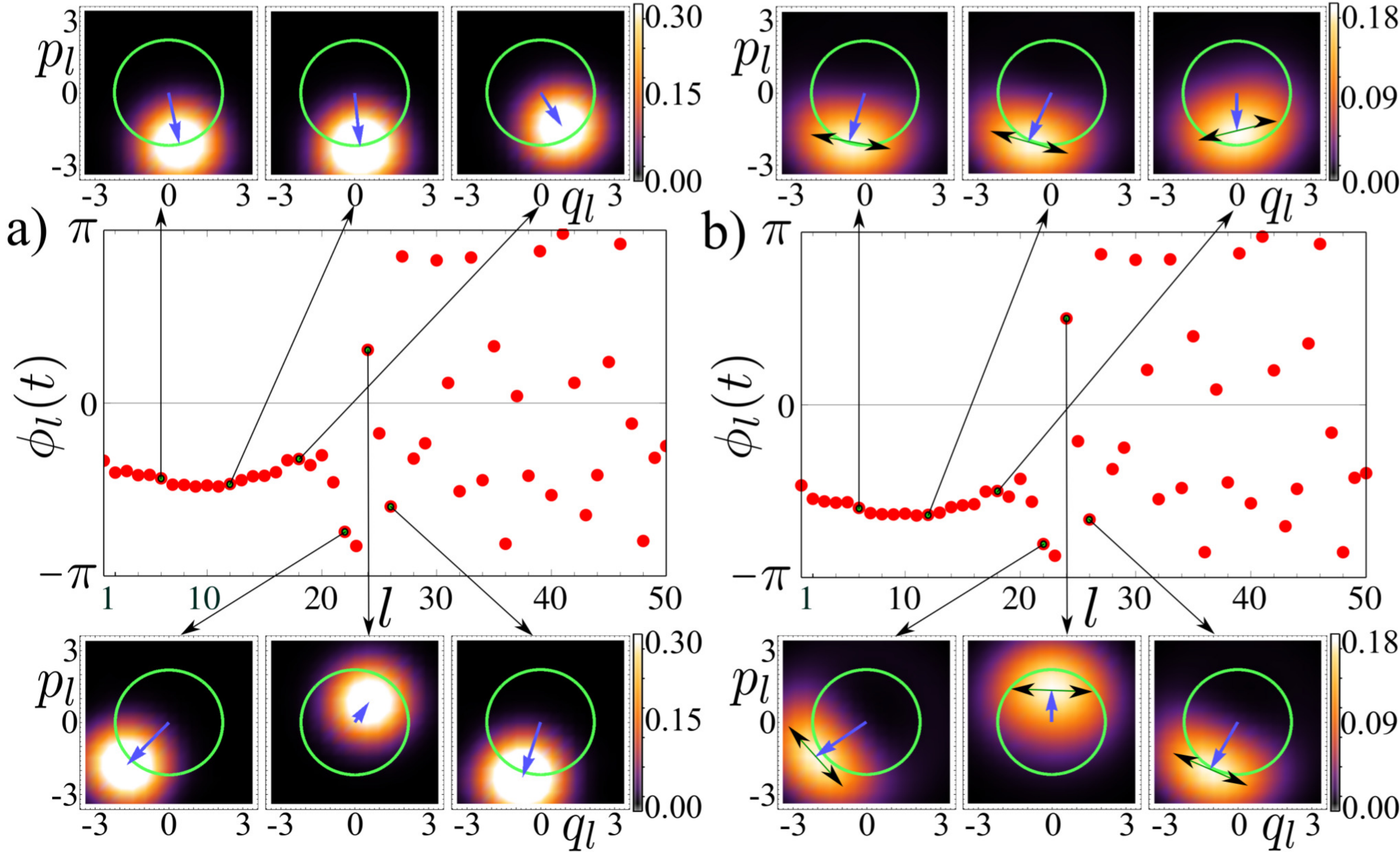}
    \caption{
    \label{Fig2}
   Quantum signatures of the classical chimera state.  (a) Snapshot of the phase chimera depicted in Fig.~\ref{Fig1} at $\kappa_{1}t_{0}=3000$. We consider an initial density matrix $\rho(t_{0})$ which is a tensor product of coherent states centered around the positions of the individual oscillators as depicted in the insets (Husimi function). (b)
   After a short-time interval $\kappa_{1}\Delta t=0.5$, quantum correlations appear in the form of squeezing (black double arrows in the insets). Parameters: $d=10$, $\kappa_2=0.2 \kappa_1$, $V=1.2\kappa_1$, and $N=50$ \cite{BAS15}.}
\end{figure}

\index{Quantum fluctuations}
As discussed in the previous section, the classical equations of motion Eq.~\eqref{QuantEqMotionNetwork} exhibit a chimera state. By using the knowledge we have about the mean field $\bm{\alpha}(t)$ in the semiclassical limit $|\alpha_{l}|\gg 1$, we can study the time evolution of the quantum fluctuations $\tilde{a}(t)$ in the co-moving frame by solving the master equation Eq.~\eqref{SemiclassMasterNetwork}. With this aim, we consider the pure coherent state as an initial density matrix $\rho(t_{0})=\bigotimes_{l=1}^{N}\ket{\alpha_{l}(t_{0})}\bra{\alpha_{l}(t_{0})}$, where $|\alpha_{l}(t_{0})|\approx 1.58$ and we choose the phases as in the left panel of Fig.~\ref{Fig2}. This initial condition corresponds to a fixed time $t_0=3000/\kappa_{1}$ in Fig.~\ref{Fig1}. In the co-moving frame, such initial condition reads $\rho_{\bm{\alpha}}(t_{0})=\bigotimes_{l=1}^{N}\ket{0_{l}}\bra{0_{l}}$.
For convenience, let us consider a representation of the bosonic operators $a_{l}=(\hat{q}_{l}+\mathrm{i}\hat{p}_{l})/\sqrt{2\hbar}$ and $\tilde{a}_{l}=(\hat{\tilde{q}}_{l}+\mathrm{i}\hat{\tilde{p}}_{l})/\sqrt{2\hbar}$ in terms of position and momentum operators
$\hat{\tilde{q}}_{l}$ and $\hat{\tilde{p}}_{l}$, respectively. We also introduce the complex variables $z_{l}=(q_{l}+\mathrm{i}p_{l})/\sqrt{2\hbar}$, $\tilde{z}_{l}=(\tilde{q}_{l}+\mathrm{i}\tilde{p}_{l})/\sqrt{2\hbar}$, which allow us to define the coordinates $\bm{z}^{T}=(z_{1},\ldots,z_{N})$ in the laboratory frame and $\bm{\tilde{z}}^{T}=(\tilde{z}_{1},\ldots,\tilde{z}_{N})$ in the co-moving frame. These coordinates are related via $\bm{z}=\bm{\alpha}(t)+\bm{\tilde{z}}$. The variables $q_{l},\tilde{q}_{l}$ and $p_{l},\tilde{p}_{l}$ denote position and conjugate momentum, respectively.

Now let us define the Wigner representation of the density operator $\rho_{\bm{\alpha}}(t)$~\cite{carmichael2009statistical}:
\begin{equation}
      \label{WignerRepresentation}
            W_{\bm{\alpha}}(\bm{\tilde{z}})=\int \frac{d^{2N}\bm{\lambda}}{\pi^{2N}} e^{-\bm{\lambda}\cdot \bm{\tilde{z}}^{\ast}+\bm{\lambda}^{\ast}\cdot \bm{\tilde{z}}}\text{tr}\left[\rho_{\bm{\alpha}}(t)e^{-\bm{\lambda}\cdot \hat{\bm{\tilde{a}}}^{\dagger}+\bm{\lambda}^{\ast}\cdot \hat{\bm{\tilde{a}}}}\right]
            \ ,
\end{equation}
where $\bm{\lambda}=(\lambda_{1},\ldots,\lambda_{N})$ denote the integration variables. 
The Husimi function $Q(\bm{z})=\frac{1}{\pi}\bra{\bm{z}}\rho(t)\ket{\bm{z}}$ is intimately related to the Wigner function via the transformation~\cite{carmichael2009statistical}
\begin{equation}
      \label{WignerHusimi}
            Q_{\bm{\alpha}}(\bm{\tilde{z}})=\frac{2}{\pi}\int W_{\bm{\alpha}}(\bm{\tilde{x}})e^{-2|\bm{\tilde{z}}-\bm{\tilde{x}}|^2}d^{2N}\bm{\tilde{x}} 
      \ .
\end{equation}
%
\index{Chimera states, quantum}
\index{Gutzwiller ansatz}
The Husimi function can be obtained numerically after solving the master equation by using the  Gutzwiller ansatz~\cite{MarquardtPhysRevLett.111.073603,SadeghpourPhysRevLett.111.234101} as we discussed in section~\ref{GutzwillerQuantumChimeraVdP}. The insets in the right panel of Fig.~\ref{Fig2} depict the Husimi functions of the individual nodes after a short evolution time $\Delta t=0.5/\kappa_1$. 
One can observe that even if one prepares the system in a separable state, quantum fluctuations arise in the form of bosonic squeezing of the oscillators~\cite{carmichael2009statistical}. In the insets of Fig.~\ref{Fig2}, the arrows indicate the direction perpendicular to the squeezing direction for the individual oscillators. For oscillators within the synchronized region, the squeezing occurs almost in the same direction. In contrast, the direction of squeezing is random for oscillators in the desynchronized region, which reflects the nature of the chimera state.

By using standard techniques of quantum optics~\cite{carmichael2009statistical}, the master equation Eq.~\eqref{SemiclassMasterNetwork} can be represented as a Fokker-Planck equation for the Wigner function Eq.~\eqref{WignerRepresentation}, which depends on the mean field solution of Eq.~\eqref{QuantEqMotionNetwork} and contains information of the chimera state
\index{Fokker-Planck equation}
%
\begin{align}
      \label{WignerFokkerPlanck}
            \frac{\partial W_{\bm{\alpha}}}{\partial t} &=\sum_{l=1}^{N}\left[2\kappa_{2}(\alpha^{\ast}_{l})^{2}\partial_{\tilde{z}^{\ast}_{l}}\tilde{z}_{l}
            +(4\kappa_2|\alpha_{l}|^{2}-\kappa_{1})\partial_{\tilde{z}_{l}}\tilde{z}_{l}+\left(2\kappa_2|\alpha_{l}|^{2}+\frac{\kappa_{1}}{2}\right)\partial^2_{\tilde{z}_{l},\tilde{z}^{\ast}_{l}}\right]W_{\bm{\alpha}}
            \nonumber \\&
            -\mathrm{i}\frac{ V}{2d}\sum_{l=1}^{N}\sum_{\substack{m=l-d\\m\neq l}}^{l+d}(\partial_{\tilde{z}^{\ast}_{m}}\tilde{z}^{\ast}_{l}-\partial_{\tilde{z}_{l}}\tilde{z}_{l})W_{\bm{\alpha}}+
            \text{H.c}
            \ .
\end{align}

For convenience, we consider the Wigner representation $W_{\bm{\alpha}}(\bm{\tilde{R}},t)$ of the density operator $\rho_{\bm{\alpha}}(t)$ in terms of the new variables $\bm{\tilde{R}}^{T}=(\tilde{q}_{1},\tilde{p}_{1},\dots,\tilde{q}_{N},\tilde{p}_{N})$. Correspondingly, the Fokker-Planck equation can be written also in terms of the quadratures $\tilde{q}_{l}$ and $\tilde{p}_{l}$
\index{Fokker-Planck equation}
%
\begin{equation}
      \label{FokkerPlanckCanonical}
          \frac{\partial W_{\bm{\alpha}}}{\partial t}=-\sum^{2N}_{i=1}\mathscr{A}_{ij}(t)\partial_{\tilde{R}_{i}}(\tilde{R}_{j}W_{\bm{\alpha}})+\frac{1}{2}\sum^{2N}_{i=1}\mathscr{B}_{ij}(t)\partial^{2}_{\tilde{R}_{i},\tilde{R}_{j}}W_{\bm{\alpha}}
      \ .
\end{equation}
Although this Fokker-Planck equation has time-dependent coefficients, one can derive an exact solution~\cite{carmichael2009statistical}
\begin{equation}
      \label{SolFokkerPlanck}
            W_{\bm{\alpha}}(\bm{\tilde{R}},t)=\frac{\exp\left(-\frac{1}{2}\bm{\tilde{R}}^{T}\cdot\mathscr{C}^{-1}(t)\cdot\bm{\tilde{R}}\right)}{(2\pi)^{N}\sqrt{\det \mathscr{C}(t)}}
       \ ,
\end{equation}
where the covariance matrix $\mathscr{C}(t)$ is a solution of the differential equation $\dot{\mathscr{C}}(t)=\mathscr{A}(t)\mathscr{C}(t)+\mathscr{C}(t)\mathscr{A}^{T}(t)+\mathscr{B}(t)$. The matrix elements 
\begin{equation}
\mathscr{C}_{ij}=\expval{\frac{1}{2}(\hat{\tilde{R}}_{i}\hat{\tilde{R}}_{j}+\hat{\tilde{R}}_{j}\hat{\tilde{R}}_{i})}_{\bm{\alpha}
}-\expval{\hat{\tilde{R}}_{i}}_{\bm{\alpha}}\expval{\hat{\tilde{R}}_{j}}_{\bm{\alpha}}
\end{equation}
of the covariance matrix contain information about the correlations between quantum fluctuations $\hat{\tilde{R}}_{2l-1}=\hat{\tilde{q}}_{l}$ and $\hat{\tilde{R}}_{2l}=\hat{\tilde{p}}_{l}$. The angular brackets $\expval{\hat{O}}_{\bm{\alpha}}=\text{tr}(\rho_{\bm{\alpha}}\hat{O})$ denote the expectation value of an operator $\hat{O}$ calculated with the density matrix $\rho_{\bm{\alpha}}$.
The solution $W_{\bm{\alpha}}(\bm{\tilde{R}},t)$ corresponds to a Gaussian distribution centered at the origin in the co-moving frame. In the laboratory frame, the Wigner function is centered at the classical trajectory $\bm{\alpha}(t)$. However, due to the chaotic nature of the classical chimera state~\cite{WOL11a}, our exact solution is just valid for short-time evolution.

\section{Chimera-like quantum correlations in the covariance matrix}
\label{CovarianceNetworkVdP}
\index{Quantum fluctuations}
\index{Chimera states, quantum}
Now let us study the consequences of the exact solution for the short-time evolution of the Wigner function. Once we obtain the solution of the equations of motion Eq.~\eqref{QuantEqMotionNetwork}, we can find the corresponding covariance matrix $\mathscr{C}(t)$. As we have defined in the introduction, a chimera state is characterized by the coexistence in space of synchronized and desynchronized dynamics. Therefore, in order to understand the quantum manifestations of a chimera state, we need to study also quantum signatures of synchronized and desynchronized dynamics.

\begin{figure}
\includegraphics[width=0.99\textwidth,clip=true]{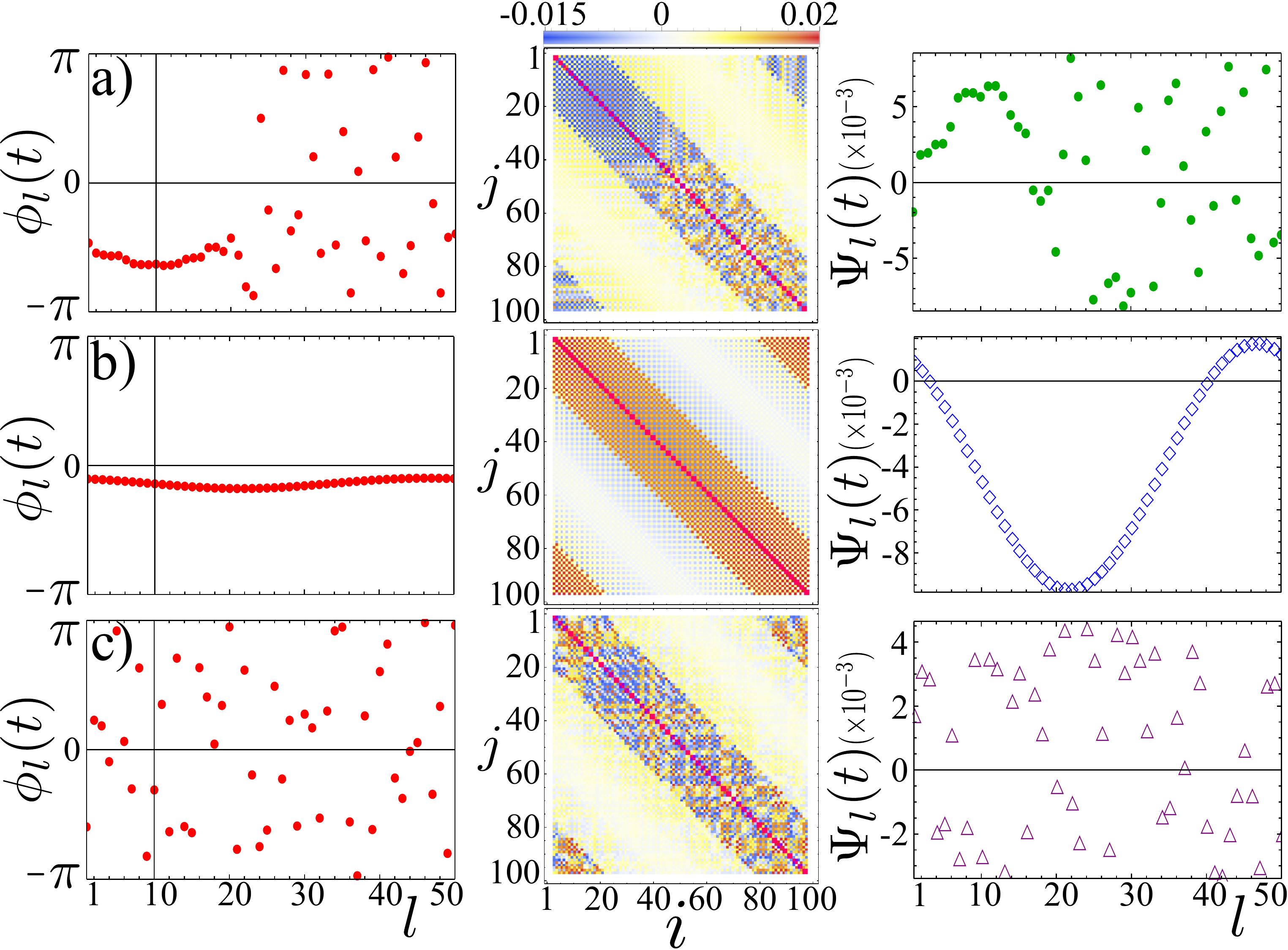}
    \caption{
    \label{Fig3}
   Quantum fluctuations after a short-time evolution. Similarly to Fig.~\ref{Fig2}, we consider an initial density matrix $\rho(t_{i})$ which is a tensor product of coherent states centered around the classical positions of the oscillators.
   Snapshots of the phase (left column) and covariance matrices (central column) after short-time evolution $\kappa_{1}\Delta t=0.5$ of the states: (a) chimera for $V=1.2\kappa_{1}$, (b) synchronized state for $V=1.6\kappa_{1}$, and (c) desynchronized state for $V=0.8\kappa_{1}$. Right column: Weighted spatial average $\Psi_{l}(t)$ of the covariance matrix for the states shown in a), b) and c), respectively. Parameters $d=10$, $\kappa_2=0.2 \kappa_1$, and $N=50$ \cite{BAS15}. 
   }
\end{figure}
%
\index{Synchronization, classical}
\index{Synchronization, quantum}

Although we consider different coupling strengths $V$, we use the same initial conditions as in Fig.~\ref{FigI1} to obtain the chimera, and completely synchronized and desynchronized states.
In order to obtain the snapshot of the chimera state depicted in Fig.~\ref{Fig3}~(a), we let the system evolve up to a time $\kappa_{1}t_{0}=3000.5$ for a coupling strength $V=1.2$. Correspondingly, to obtain the snapshot of the synchronized solution shown in Fig.~\ref{Fig3}~(b), we let the system evolve a time
$\kappa_{1}t_{\text{Syn}}=25.5$ for $V=1.6$. Finally, the snapshot of the desynchronized state in Fig.~\ref{Fig3}~(c) is obtained after a time evolution $\kappa_{1}t_{\text{desyn}}=8000.5$ for $V=0.8$. The left column of Fig.~\ref{Fig3} show snapshots of the phases for (a) chimera, (b) completely synchronized, and (c) completely desynchronized mean-field solutions of Eq.~\eqref{QuantEqMotionNetwork}. The central column of Fig.~\ref{Fig3} depicts the corresponding covariance matrices after a short evolution time $\Delta t=0.5/\kappa_{1}$.
For every plot, we have initialized the system at time $t_{i}$ as a tensor product of coherent states $\ket{\alpha_{l}(t_{i})}$ centered at the positions $\alpha_{l}(t_{i})$ of the individual oscillators. As a consequence, the covariance matrix at the initial time is diagonal $\mathscr{C}_{2l-1, 2l-1}(t_i)=\expval{\hat{\tilde{q}}^{2}_{l}}_{\bm{\alpha}}=\hbar/2$ and $\mathscr{C}_{2l, 2l}(t_i)=\expval{\hat{\tilde{p}}^{2}_{l}}_{\bm{\alpha}}=\hbar/2$, which reflects the Heisenberg uncertainty principle because $\expval{\hat{\tilde{q}}_{l}}_{\bm{\alpha}}=\expval{\hat{\tilde{p}}_{l}}_{\bm{\alpha}}=0$.

After a short evolution time, quantum correlations are built up due to the coupling between the oscillators, and the covariance matrix exhibits a nontrivial structure which is influenced by the mean field solution. For example, the central panel of Fig.~\ref{Fig3}~(a) shows a matrix plot of the covariance matrix corresponding to a chimera state obtained from the same initial condition as in Fig.~\ref{Fig2}. The covariance matrix acquires a block structure, where the upper $40\times 40 $ block (corresponding to nodes $l=1,\dots,20$) shows a regular pattern matching the synchronized region of the chimera state. Similarly, the lower $60\times 60$ block shows an irregular structure which corresponds to the desynchronized dynamics of the oscillators $l=21,\dots,50$. In a similar fashion, Figs.~\ref{Fig3}~(b) and (c) show the matrix $\mathscr{C}$ for completely synchronized and desynchronized states, respectively.
In the case of a chimera state, this coincides with the results shown in Fig.~\ref{Fig2}, where the squeezing direction of the oscillators is related to the classical solution. In order to quantify these observations we define the weighted  correlation as
\begin{equation}
      \label{KuramotoOrderPar}
            \Psi_{l}(t)=\frac{V}{2d}\sum^{l+d}_{\substack{m=l-d\\m\neq l}}\mathscr{C}_{2l, 2m}(t)
      \ .
\end{equation}
This spatial average 
highlights the structure of the covariance matrix. The right column of Fig.~\ref{Fig3} shows 
$\Psi_{l}(t)$ for (a) chimera, (b) synchronized, and (c) desynchronized states. The chimera state exhibits a regular and an irregular domain, exactly as the classical chimera does. 

\section{Quantum mutual information and chimera states}
\label{QuantMutualInfoNetworkVdP}
\index{Quantum mutual information}
\index{Covariance matrix}
\index{Chimera states, classical}
\index{Chimera states, quantum}
Now let us consider a partition of the network into spatial domains of size $L$  and $N-L$, which we call \textit{Alice} (A) and \textit{Bob} (B), respectively. This partition can be represented by considering a decomposition of the covariance matrix
\begin{equation}
       \label{CovMatrix}
       \mathscr{C}(t) =\left(%
\begin{array}{ccc}
\mathscr{C}_{\text{A}}(t)  &  \mathscr{C}_{\text{AB}}(t) \\
 \mathscr{C}^{T}_{\text{AB}}(t) & \mathscr{C}_{\text{B}}(t)
\end{array}
\right)
\end{equation}
\begin{figure}
\includegraphics[width=0.99\textwidth,clip=true]{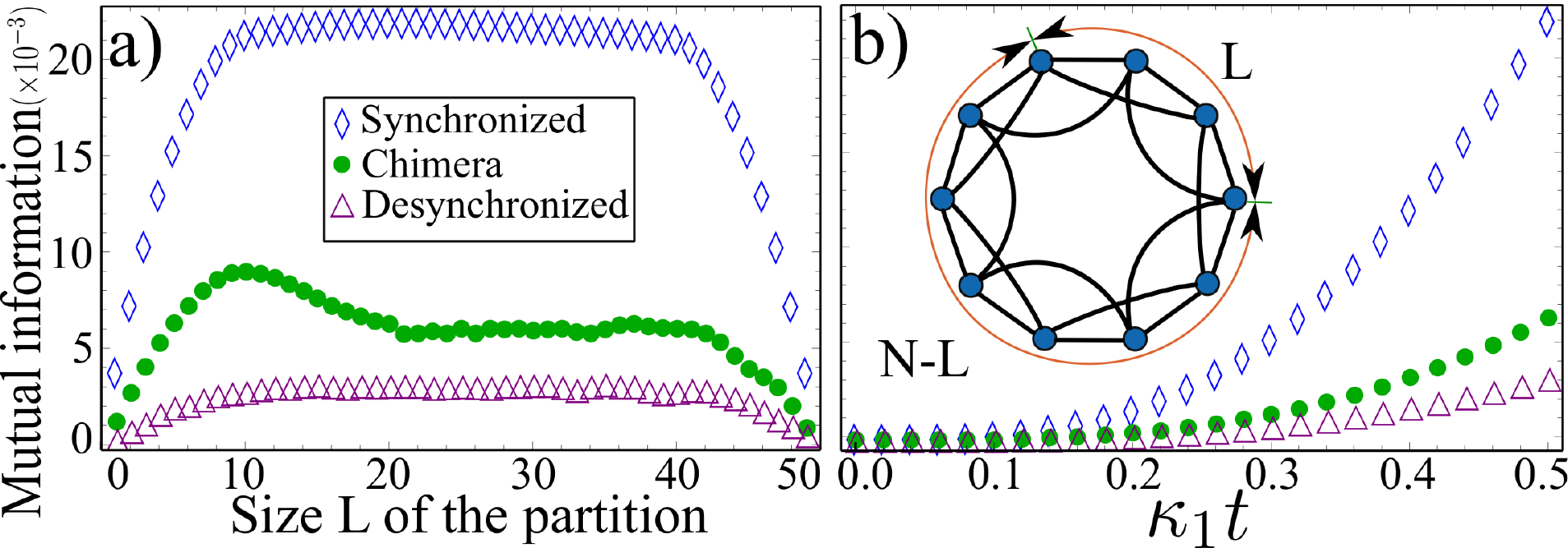}
    \caption{
    \label{Fig4}
   R{\'e}nyi quantum mutual information for the states shown in Fig.~\ref{Fig3}. The green dots, blue diamonds, and purple triangles represent the chimera, synchronized, and desynchronized states, respectively. a) Gaussian R{\'e}nyi-2  mutual information $\mathcal{I}_{2}(\rho_{\text{A:B}})$ as a function of the size $L$ of \textit{Alice} after an evolution time $\Delta t=0.5/\kappa_{1}$. b) The time evolution of the mutual information during the time interval $\Delta t$ for a fixed size $L_{c}=20$. Inset: scheme of the nonlocally coupled network. Parameters: $d=10$, $\kappa_2=0.2\kappa_1$, and $N=50$  \cite{BAS15}.}
\end{figure}
%
\index{R{\'e}nyi entropy}
To study the  interplay between synchronized and desynchronized dynamics, which is characteristic of a chimera state, we propose the use of an entropy measure~\cite{VanLoockRevModPhys.77.513,LloydRevModPhys.84.621,FazioPhysRevA.91.012301}. Of particular interest is the R{\'e}nyi entropy $S_{\mu}(\rho)=(1-\mu)^{-1}\ln\text{tr}(\rho^{\mu})$, $\mu \in \mathbb{N}$, of the density matrix $\rho$, which is discussed in Ref.~\cite{AdessoPhysRevLett.109.190502}. In terms of the Wigner representation of $\rho_{\bm{\alpha}}$, the R{\'e}nyi entropy for $\mu=2$ reads $S_{2}(\rho_{\bm{\alpha}})=-\ln\left[\int W^{2}_{\bm{\alpha}}(\bm{\tilde{R}},t) d^{2N}\bm{\tilde{R}} \right]$. Now let us consider the bipartite Gaussian state $\rho_{\text{AB}}=\rho_{\bm{\alpha}}$ composed of \textit{Alice}
and  \textit{Bob} and define the tensor product $\rho_{\text{Ref}}=\rho_{\text{A}}\otimes\rho_{\text{B}}$ of the two marginals $\rho_{\text{A}}$ and $\rho_{\text{B}}$.

To measure Gaussian R{\'e}nyi-2  mutual information $\mathcal{I}_{2}(\rho_{\text{A:B}})=S_{2}(\rho_{\text{A}})+S_{2}(\rho_{\text{B}})-S_{2}(\rho_{\text{AB}})$, we require the calculation of the relative sampling entropy between the total density matrix $\rho_{\text{AB}}$ and the reference state $\rho_{\text{Ref}}$ as shown in Ref.~\cite{AdessoPhysRevLett.109.190502}. This leads to a formula  $\mathcal{I}_{2}(\rho_{\text{A:B}})=\frac{1}{2}\ln\left(\det\mathscr{C}_{A}\det\mathscr{C}_{B}/\det\mathscr{C}\right)$ in terms of the covariance matrix Eq.~\eqref{CovMatrix}.
Figure~\ref{Fig4}~(a) shows the variation of $\mathcal{I}_{2}(\rho_{\text{A:B}})$ as a function of the size $L$ of the partition after an evolution time $\Delta t=0.5/\kappa_{1}$. One can observe that for a chimera state the mutual information is asymmetric as a function of $L$ and there is a critical size $L_{c}=20$, where a dramatic change of the correlations occurs.

Now let us consider the chimera state shown in Fig.~\ref{Fig2}, and let us consider a partition where the size of \textit{Alice} is $L_{c}=20$. Fig.~\ref{Fig4}~(b) shows the time evolution of mutual information for such a state. In addition, by using the same partition as for the chimera state,
we calculate the mutual information for the synchronized and desynchronized states depicted in Figs.~\ref{Fig3}~(b) and (c), respectively. Our results reveal that the chimera state has a mutual information which lies between the values for synchronized and desynchronized states. This resembles the definition of a chimera state given at the beginning of the article.

\section{Summary and outlook}
\label{sec discussion}
%
\index{R{\'e}nyi entropy}
\index{Quantum mutual information}
We have shown that quantum signatures of chimera states appear in the covariance matrix and in measures of mutual information. To quantify the structure of the covariance matrix, we have introduced a weighted spatial average of the quantum correlation, which reveals the nature of the classical trajectory, i.e., chimera, completely synchronized, or completely desynchronized state. The mutual information for a bipartite state $\mathcal{I}_{2}(\rho_{\text{A:B}})$ extends the definition of a chimera to the quantum regime and highlights the relation to quantum information theory. A possible experimental realization of our model could be carried out by means of trapped ions~\cite{WinelandRevModPhys.75.281}, as it was suggested in Ref.~\cite{SadeghpourPhysRevLett.111.234101}. Other experimental possibilities include  SQUID metamaterials~\cite{lazarides2015chimeras} and Bose-Einstein condensation in the presence of dissipation and external driving~\cite{DiehlPhysRevLett.110.195301,DiehlPhysRevX.4.021010,
kasprzak2006bose}. In this context our approach is particularly interesting, because the continuum limit of the mean field Eq.~\eqref{QuantEqMotionNetwork} is a complex Ginzburg-Landau 
equation, which is nonlocal in space~\cite{KUR02a}. In this sense, our linearized master equation Eq.~\eqref{SemiclassMasterNetwork} enables us to study the Bogoliubov excitations above the mean field solution.

\section*{Acknowledgments}

V.M. Bastidas thanks L. M. Valencia and Y. Sato. The authors acknowledge inspiring discussions with J. Cerrillo, S. Restrepo, G. Schaller, and P. Strasberg. This work was supported by DFG in the framework of SFB 910.


\printindex
\appendix

\section{Appendix}
\label{appendix}
\index{Quantum fluctuations}
\index{Semiclassical dynamics}
%
In this appendix we discuss the Gaussian quantum fluctuations of the quartic harmonic oscillator
\begin{equation}
      \label{HamQuartOsci}
            H=\frac{p^2}{2m}+\frac{m\omega q^2}{2}+\frac{\lambda}{4}q^4=\omega \left(a^{\dagger}a+\frac{1}{2}\right)+\lambda\left(\frac{1}{4m\omega}\right)^2(a^{\dagger}+a)^4
       \ .
\end{equation}
We now consider  the time-dependent displacement operator~\cite{GlauberPhysRev.177.1857}
\begin{align}
     \label{ProdDisOp}
      \mathcal{D}\left[\alpha(t)\right]&=\exp\left[\alpha(t)\hat{a}^{\dagger}-\alpha^{*}(t) \hat{a}\right]
      \nonumber\\
      &=\exp\left[-\frac{|\alpha(t)|^{2}}{2}
      \right]\exp\left[\alpha(t)\hat{a}^{\dagger}\right]\exp\left[-\alpha^{*}(t) \hat{a}\right]
	\ .
\end{align}
Under this Gauge transformation, the Schr\"odinger equation $\mathrm{i}\partial_{t}|\Psi(t)\rangle=H|\Psi(t)\rangle$ is transformed to $\mathrm{i}\partial_{t}|\Psi_{\alpha}(t)\rangle=H^{(\alpha)}(t)|\Psi_{\alpha}(t)\rangle$, where
$|\Psi_{\alpha}(t)\rangle=\mathcal{D}^{\dagger}\left[\alpha(t)\right]|\Psi(t)\rangle$ and
\begin{equation}
     \label{QuantumHarmAppr}
      \hat{H}^{(\alpha)}(t)=\mathcal{D}^{\dagger}\left[\alpha(t)\right](H-\mathrm{i}\partial_t)\mathcal{D}\left[\alpha(t)\right]
	  \ .
\end{equation}
For later purposes, we need to use the identity
\begin{equation}
      \label{DerDisp}
          \mathrm{i}\mathcal{D}^{\dagger}\left[\alpha(t)\right]\partial_t\mathcal{D}\left[\alpha(t)\right]
          =\frac{\mathrm{i}}{2}[\dot{\alpha}(t)\alpha^{*}(t)-\alpha(t)\dot{\alpha}^{*}(t)]+\mathrm{i}[\dot{\alpha}(t)a^{\dagger}-\dot{\alpha}^{*}(t)a]
      \ .
\end{equation}
After some algebraic manipulations we can write 
\begin{align}
      \label{QuartHamExp}
            \hat{H}^{(\alpha)}(t)&=\frac{\lambda}{16}\left(\frac{1}{m\omega}\right)^2(a^{\dagger}+a)^4+\frac{1}{2}\left(\frac{1}{m\omega}\right)^2(a^{\dagger}+a)^3\text{Re}[\alpha(t)]
            \nonumber\\&
            +\omega a^{\dagger}a+\frac{3\lambda}{2}\left(\frac{1}{m\omega}\right)^2(a^{\dagger}+a)^2(\text{Re}[\alpha(t)])^{2}
            \nonumber\\&
            -\mathrm{i}[\dot{\alpha}(t)a^{\dagger}-\dot{\alpha}^{*}(t)a]+\omega(\alpha^{*}a+\alpha a^{\dagger})+2\lambda\left(\frac{1}{m\omega}\right)^2(a^{\dagger}+a)(\text{Re}[\alpha(t)])^{3}
            \nonumber\\&
            -\frac{\mathrm{i}}{2}[\dot{\alpha}(t)\alpha^{*}(t)-\alpha(t)\dot{\alpha}^{*}(t)]+\omega|\alpha(t)|^2+\lambda\left(\frac{1}{m\omega}\right)^2(\text{Re}[\alpha(t)])^{4}
       \ .
\end{align}
To study the quantum fluctuations about a semiclassical trajectory, we assume that $|\alpha(t)|\gg 1$. To obtain the quadratic fluctuations we must neglect the non-Gaussian terms in Eq.~\eqref{QuartHamExp}. In addition, to choose the trajectory we have the condition 
\begin{equation}
      \label{QuartHamClass}
            \dot{\alpha}(t)=-\mathrm{i}\left(\omega\alpha(t) +2\lambda\left(\frac{1}{m\omega}\right)^2(\text{Re}[\alpha(t)])^{3}\right)
       \ ,
\end{equation}
which corresponds to the equations of motion. We can go a step further and define the classical Hamiltonian function
\begin{align}
      \label{ClassicalQuarticHamFunc}
            H(\alpha,\alpha^*)&=\omega|\alpha(t)|^2+\lambda\left(\frac{1}{m\omega}\right)^2(\text{Re}[\alpha(t)])^{4}
            \nonumber\\&
            =\omega \alpha^{*}(t)\alpha(t)+\lambda\left(\frac{1}{m\omega}\right)^2\left[\frac{\alpha^{*}(t)+\alpha(t)}{2}\right]^{4}
      \ ,
\end{align}
from which we obtain the equations of motion Eq.~\eqref{QuartHamClass}. In so doing we define the Poisson bracket
$\{F(\alpha,\alpha^*),G(\alpha,\alpha^*)\}=-\mathrm{i}(\partial_{\alpha}F\partial_{\alpha^{*}}G-\partial_{\alpha}G\partial_{\alpha^{*}}F)$. By using this definition we obtain the equations of motion Eq.~\eqref{QuartHamClass} as $\dot{\alpha}(t)=-\mathrm{i}\partial_{\alpha^{*}}H(\alpha,\alpha^*)$.

Now it is clear the role of each one of the terms in Eq.~\eqref{QuartHamExp}.  In contrast, the quadratic terms give us the first quantum corrections about the semiclassical trajectory. To study these quantum fluctuations we need to study the quadratic Hamiltonian
\begin{equation}
      \label{GaussianFluctQuartOsc}
            \hat{H}_{\text{Q}}^{(\alpha)}(t)=\omega a^{\dagger}a+\frac{3\lambda}{2}\left(\frac{1}{m\omega}\right)^2(a^{\dagger}+a)^2(\text{Re}[\alpha(t)])^{2}+L(\alpha,\alpha^*)
      \ ,
\end{equation}
where $L(\alpha,\alpha^*)=-\frac{\mathrm{i}}{2}[\dot{\alpha}(t)\alpha^{*}(t)-\alpha(t)\dot{\alpha}^{*}(t)]+H(\alpha,\alpha^*)$ is the Lagrangian.


\begin{thebibliography}{10}
\providecommand{\url}[1]{{#1}}
\providecommand{\urlprefix}{URL }
\expandafter\ifx\csname urlstyle\endcsname\relax
  \providecommand{\doi}[1]{DOI \discretionary{}{}{}#1}\else
  \providecommand{\doi}{DOI \discretionary{}{}{}\begingroup
  \urlstyle{rm}\Url}\fi

\bibitem{PAN15}
M.J. Panaggio, D.M. Abrams, Nonlinearity \textbf{28}, R67 (2015)

\bibitem{KUR02a}
Y.~Kuramoto, D.~Battogtokh, Nonlin. Phen. in Complex Sys. \textbf{5}(4), 380
  (2002)

\bibitem{ABR04}
D.M. Abrams, S.H. Strogatz, Phys.~Rev.~Lett. \textbf{93}(17), 174102 (2004)

\bibitem{LAI09}
C.R. Laing, Physica D \textbf{238}(16), 1569 (2009)

\bibitem{MOT10}
A.E. Motter, Nature Physics \textbf{6}(3), 164 (2010)

\bibitem{OME10a}
O.E. Omel'chenko, M.~Wolfrum, Y.~Maistrenko, Phys. Rev.~E \textbf{81}(6),
  065201(R) (2010)

\bibitem{OME12a}
O.E. Omel'chenko, M.~Wolfrum, S.~Yanchuk, Y.~Maistrenko, O.~Sudakov, Phys.
  Rev.~E \textbf{85}, 036210 (2012)

\bibitem{MAR10b}
E.A. Martens, Chaos \textbf{20}(4), 043122 (2010)

\bibitem{WOL11a}
M.~Wolfrum, O.E. Omel'chenko, S.~Yanchuk, Y.~Maistrenko, Chaos \textbf{21},
  013112 (2011)

\bibitem{BOU14}
T.~Bountis, V.~Kanas, J.~Hizanidis, A.~Bezerianos, Eur. Phys.~J. Special Topic
  \textbf{223}(4), 721 (2014)

\bibitem{LAI10}
C.R. Laing, Phys. Rev. E \textbf{81}(6), 066221 (2010)

\bibitem{OME15}
I.~Omelchenko, A.~Provata, J.~Hizanidis, E.~Sch{\"o}ll, P.~H{\"o}vel, Phys.
  Rev. E \textbf{91}, 022917 (2015)

\bibitem{OME11}
I.~Omelchenko, Y.~Maistrenko, P.~H{\"o}vel, E.~Sch{\"o}ll, Phys. Rev. Lett.
  \textbf{106}, 234102 (2011)

\bibitem{OME12}
I.~Omelchenko, B.~Riemenschneider, P.~H{\"o}vel, Y.~Maistrenko, E.~Sch{\"o}ll,
  Phys. Rev.~E \textbf{85}, 026212 (2012)

\bibitem{OME13}
I.~Omelchenko, O.E. Omel'chenko, P.~H{\"o}vel, E.~Sch{\"o}ll, Phys. Rev. Lett.
  \textbf{110}, 224101 (2013)

\bibitem{HIZ13}
J.~Hizanidis, V.~Kanas, A.~Bezerianos, T.~Bountis, Int. J. Bifurcation Chaos
  \textbf{24}(03), 1450030 (2014)

\bibitem{VUE14a}
A.~V{\"u}llings, J.~Hizanidis, I.~Omelchenko, P.~H{\"o}vel, New J.~Phys.
  \textbf{16}, 123039 (2014)

\bibitem{OME15a}
I.~Omelchenko, A.~Zakharova, P.~H{\"o}vel, J.~Siebert, E.~Sch{\"o}ll, arXiv p.
  1503.03377 (2015)

\bibitem{ROS14a}
D.P. Rosin, D.~Rontani, N.D. Haynes, E.~Sch{\"o}ll, D.J. Gauthier, Phys. Rev.~E
  \textbf{90}, 030902(R) (2014)

\bibitem{SHI04}
S.i. Shima, Y.~Kuramoto, Phys. Rev.~E \textbf{69}(3), 036213 (2004)

\bibitem{MAR10}
E.A. Martens, C.R. Laing, S.H. Strogatz, Phys. Rev. Lett. \textbf{104}(4),
  044101 (2010)

\bibitem{PAN13}
M.J. Panaggio, D.M. Abrams, Phys. Rev. Lett. \textbf{110}, 094102 (2013)

\bibitem{PAN15a}
M.J. Panaggio, D.M. Abrams, Phys. Rev. E \textbf{91}, 022909 (2015)

\bibitem{SET08}
G.C. Sethia, A.~Sen, F.M. Atay, Phys.~Rev.~Lett. \textbf{100}(14), 144102
  (2008)

\bibitem{MAI14}
Y.~Maistrenko, A.~Vasylenko, O.~Sudakov, R.~Levchenko, V.L. Maistrenko, arXiv:
  1402.1363v1  (2014)

\bibitem{XIE14}
J.~Xie, E.~Knobloch, H.C. Kao, Phys. Rev. E \textbf{90}, 022919 (2014)

\bibitem{SET13}
G.C. Sethia, A.~Sen, G.L. Johnston, Phys. Rev. E \textbf{88}(4), 042917 (2013)

\bibitem{SET14}
G.C. Sethia, A.~Sen, Phys. Rev. Lett. \textbf{112}, 144101 (2014)

\bibitem{ZAK14}
A.~Zakharova, M.~Kapeller, E.~Sch{\"o}ll, Phys.~Rev.~Lett. \textbf{112}, 154101
  (2014)

\bibitem{YEL14}
A.~Yeldesbay, A.~Pikovsky, M.~Rosenblum, Phys. Rev. Lett. \textbf{112}, 144103
  (2014)

\bibitem{SCH14g}
L.~Schmidt, K.~Krischer, arXiv preprint arXiv:1409.1479  (2014)

\bibitem{BOE15}
F.~B{\"o}hm, A.~Zakharova, E.~Sch{\"o}ll, K.~L{\"u}dge, Phys. Rev. E
  \textbf{91}(4), 040901 (R) (2015)

\bibitem{KO08}
T.W. Ko, G.B. Ermentrout, Phys. Rev.~E \textbf{78}, 016203 (2008)

\bibitem{SHA10}
M.~Shanahan, Chaos \textbf{20}(1), 013108 (2010)

\bibitem{LAI12}
C.R. Laing, K.~Rajendran, I.G. Kevrekidis, Chaos \textbf{22}(1), 013132 (2012)

\bibitem{YAO13}
N.~Yao, Z.G. Huang, Y.C. Lai, Z.~Zheng, Scientific Reports \textbf{3}, 3522
  (2013)

\bibitem{ZHU14}
Y.~Zhu, Z.~Zheng, J.~Yang, Phys. Rev. E \textbf{89}, 022914 (2014)

\bibitem{BUS15}
A.~Buscarino, M.~Frasca, L.V. Gambuzza, P.~H{\"o}vel, Phys. Rev.~E
  \textbf{91}(2), 022817 (2015)

\bibitem{RAT00}
N.C. Rattenborg, C.J. Amlaner, S.L. Lima, Neurosci. Biobehav. Rev. \textbf{24},
  817 (2000)

\bibitem{LAI01}
C.R. Laing, C.C. Chow, Neural Computation \textbf{13}(7), 1473 (2001)

\bibitem{SAK06a}
H.~Sakaguchi, Phys. Rev.~E \textbf{73}(3), 031907 (2006)

\bibitem{ROT14}
A.~Rothkegel, K.~Lehnertz, New J. of Phys. \textbf{16}, 055006 (2014)

\bibitem{FIL08}
A.E. Filatova, A.E. Hramov, A.A. Koronovskii, S.~Boccaletti, Chaos \textbf{18},
  023133 (2008)

\bibitem{GON14}
J.C. Gonzalez-Avella, M.G. Cosenza, M.S. Miguel, Physica~A \textbf{399}(0), 24
  (2014)

\bibitem{OME13a}
O.E. Omel'chenko, Nonlinearity \textbf{26}(9), 2469 (2013)

\bibitem{SIE14c}
J.~Sieber, O.E. Omel'chenko, M.~Wolfrum, Phys. Rev. Lett. \textbf{112}, 054102
  (2014).
\newblock \doi{10.1103/physrevlett.112.054102}

\bibitem{BIC14}
C.~Bick, E.A. Martens, arXiv \textbf{1402.6363v1} (2014)

\bibitem{HAG12}
A.M. Hagerstrom, T.E. Murphy, R.~Roy, P.~H{\"o}vel, I.~Omelchenko,
  E.~Sch{\"o}ll, Nature Physics \textbf{8}, 658 (2012)

\bibitem{TIN12}
M.R. Tinsley, S.~Nkomo, K.~Showalter, Nature Physics \textbf{8}, 662 (2012)

\bibitem{NKO13}
S.~Nkomo, M.R. Tinsley, K.~Showalter, Phys. Rev. Lett. \textbf{110}, 244102
  (2013)

\bibitem{MAR13}
E.A. Martens, S.~Thutupalli, A.~Fourri{\`e}re, O.~Hallatschek, Proc. Nat. Acad.
  Sciences \textbf{110}, 10563 (2013)

\bibitem{LAR13}
L.~Larger, B.~Penkovsky, Y.~Maistrenko, Phys. Rev. Lett. \textbf{111}, 054103
  (2013)

\bibitem{GAM14}
L.V. Gambuzza, A.~Buscarino, S.~Chessari, L.~Fortuna, R.~Meucci, M.~Frasca,
  Phys. Rev. E \textbf{90}, 032905 (2014)

\bibitem{SCH14a}
L.~Schmidt, K.~Sch{\"o}nleber, K.~Krischer, V.~Garcia-Morales, Chaos
  \textbf{24}(1), 013102 (2014)

\bibitem{WIC13}
M.~Wickramasinghe, I.Z. Kiss, PLoS ONE \textbf{8}(11), e80586 (2013)

\bibitem{VIK14}
E.A. Viktorov, T.~Habruseva, S.P. Hegarty, G.~Huyet, B.~Kelleher, Phys. Rev.
  Lett. \textbf{112}, 224101 (2014)

\bibitem{LAZ15}
N.~Lazarides, G.~Neofotistos, G.~Tsironis, Phys.~Rev.~B \textbf{91}, 054303
  (2015)

\bibitem{KAP12}
M.~Kapitaniak, K.~Czolczynski, P.~Perlikowski, A.~Stefanski, T.~Kapitaniak,
  Phys. Rep. \textbf{517}, 1 (2012)

\bibitem{SadeghpourPhysRevLett.111.234101}
T.E. Lee, H.R. Sadeghpour, Phys. Rev. Lett. \textbf{111}, 234101 (2013)

\bibitem{BruderPhysRevLett.112.094102}
S.~Walter, A.~Nunnenkamp, C.~Bruder, Phys. Rev. Lett. \textbf{112}, 094102
  (2014)

\bibitem{MarquardtPhysRevLett.111.073603}
M.~Ludwig, F.~Marquardt, Phys. Rev. Lett. \textbf{111}, 073603 (2013)

\bibitem{PachonKuramoto}
I.~{Hermoso de Mendoza}, L.A. Pach{\'o}n, J.~G{\'o}mez-Garde{\~n}es, D.~Zueco,
  Phys. Rev. E \textbf{90}, 052904 (2014)

\bibitem{FazioPhysRevLett.111.103605}
A.~Mari, A.~Farace, N.~Didier, V.~Giovannetti, R.~Fazio, Phys. Rev. Lett.
  \textbf{111}, 103605 (2013)

\bibitem{LloydRevModPhys.84.621}
C.~Weedbrook, S.~Pirandola, R.~Garc{\'i}a-Patr{\'o}n, N.J. Cerf, T.C. Ralph,
  J.H. Shapiro, S.~Lloyd, Rev. Mod. Phys. \textbf{84}, 621 (2012)

\bibitem{VanLoockRevModPhys.77.513}
S.L. Braunstein, P.~van Loock, Rev. Mod. Phys. \textbf{77}, 513 (2005)

\bibitem{Zambrini}
{Manzano Gonzalo}, {Galve Fernando}, {Giorgi Gian Luca},
  {Hern{\'a}ndez-Garc{\'i}a Emilio}, {Zambrini Roberta}, Sci. Rep. \textbf{3}
  (2013)

\bibitem{ZambriniOsc}
G.~Manzano, F.~Galve, R.~Zambrini, Phys. Rev. A \textbf{87}, 032114 (2013)

\bibitem{LeeWang}
T.E. Lee, C.K. Chan, S.~Wang, Phys. Rev. E \textbf{89}, 022913 (2014)

\bibitem{ZambriniSpin}
G.L. Giorgi, F.~Plastina, G.~Francica, R.~Zambrini, Phys. Rev. A \textbf{88},
  042115 (2013)

\bibitem{FazioPhysRevA.91.012301}
V.~Ameri, M.~Eghbali-Arani, A.~Mari, A.~Farace, F.~Kheirandish, V.~Giovannetti,
  R.~Fazio, Phys. Rev. A \textbf{91}, 012301 (2015)

\bibitem{BAS15}
V.~Bastidas, I.~Omelchenko, A.~Zakharova, E.~Sch{\"o}ll, T.~Brandes, arXiv
  preprint arXiv:1505.02639  (2015)

\bibitem{Viennot}
D.~Viennot, L.~Aubourg, arXiv preprint arXiv:1408.4585  (2014)

\bibitem{DiehlPhysRevLett.110.195301}
L.M. Sieberer, S.D. Huber, E.~Altman, S.~Diehl, Phys. Rev. Lett. \textbf{110},
  195301 (2013)

\bibitem{DiehlPhysRevX.4.021010}
U.C. T{\"a}uber, S.~Diehl, Phys. Rev. X \textbf{4}, 021010 (2014)

\bibitem{AdessoPhysRevLett.109.190502}
G.~Adesso, D.~Girolami, A.~Serafini, Phys. Rev. Lett. \textbf{109}, 190502
  (2012)

\bibitem{van1920theory}
B.~{Van der Pol}, Radio Review \textbf{1}(1920), 701 (1920)

\bibitem{WinelandRevModPhys.75.281}
D.~Leibfried, R.~Blatt, C.~Monroe, D.~Wineland, Rev. Mod. Phys. \textbf{75},
  281 (2003)

\bibitem{ArmouPhysRevLett.104.053601}
D.A. Rodrigues, A.D. Armour, Phys. Rev. Lett. \textbf{104}, 053601 (2010)

\bibitem{HammererPhysRevX.4.011015}
N.~L{\"o}rch, J.~Qian, A.~Clerk, F.~Marquardt, K.~Hammerer, Phys. Rev. X
  \textbf{4}, 011015 (2014)

\bibitem{carmichael2009statistical}
H.J. Carmichael, \emph{{Statistical Methods in Quantum Optics 2: Non-Classical
  Fields}} (Springer Science \& Business Media, 2009)

\bibitem{GlauberPhysRev.177.1857}
K.E. Cahill, R.J. Glauber, Phys. Rev. \textbf{177}, 1857 (1969)

\bibitem{HaakePhysRevLett.108.073601}
A.~Altland, F.~Haake, Phys. Rev. Lett. \textbf{108}, 073601 (2012)

\bibitem{lazarides2015chimeras}
N.~Lazarides, G.~Neofotistos, G.~Tsironis, Physical Review B \textbf{91}(5),
  054303 (2015)

\bibitem{kasprzak2006bose}
J.~Kasprzak, M.~Richard, S.~Kundermann, A.~Baas, P.~Jeambrun, J.~Keeling,
  F.~Marchetti, M.~Szyma{\'n}ska, R.~Andre, J.~Staehli, et~al., Nature
  \textbf{443}(7110), 409 (2006)

\end{thebibliography}
\end{document}